\documentclass[amsmath,amssymb,11pt,superscriptaddress
,nofootinbib]{revtex4}

\usepackage{graphicx}
\usepackage{bm}

\newcommand{\dd}{{\rm d}}

\begin{document}
\baselineskip5.5mm

\thispagestyle{empty}

{\baselineskip0pt
\leftline{\baselineskip14pt\sl\vbox to0pt{
               \hbox{\it Yukawa Institute for Theoretical Physics} 
             \hbox{\it Kyoto University}
               \vss}}
\rightline{\baselineskip16pt\rm\vbox to20pt{
            {
            \hbox{YITP-11-41}
            }
\vss}}%
}

\author{Chul-Moon Yoo}\email{yoo@yukawa.kyoto-u.ac.jp}
\affiliation{
Yukawa Institute for Theoretical Physics, Kyoto University
Kyoto 606-8502, Japan
}

\author{Sugure Tanzawa}\email{tanzawa@yukawa.kyoto-u.ac.jp}
\affiliation{
Yukawa Institute for Theoretical Physics, Kyoto University
Kyoto 606-8502, Japan
}

\author{Misao Sasaki}\email{misao@yukawa.kyoto-u.ac.jp}
\affiliation{
Yukawa Institute for Theoretical Physics, Kyoto University
Kyoto 606-8502, Japan
}
\affiliation{
Korea Institute for Advanced Study
207-43 Cheongnyangni 2-dong, Dongdaemun-gu, 
Seoul 130-722, Republic of Korea
}

\vskip2cm
\title{Gregory-Laflamme instability of a slowly rotating black string
}

\begin{abstract}
We study the Gregory-Laflamme instability 
of a 5-dimensional slowly rotating black string 
in which the 4-dimensional section is 
described by the Kerr black hole. 
We treat the rotation in a perturbative way 
introducing a small parameter for the rotation.  
It is found that rotation makes 
the Gregory-Laflamme instability stronger. 
Both the critical wavelength at the onset of instability 
and the growth time-scale are found to decrease
as the rotation increases. 
\end{abstract}

\maketitle
\pagebreak

\section{Introduction}
\label{sec:intro}

The Gregory-Laflamme(GL) instability is an $s$-wave instability
for black $p$-brane solutions~\cite{Gregory:1993vy,Gregory:1994bj}. 
This instability is often quoted 
in the context of gravitational instabilities of black objects. 
For instance, the gravitational instability of 
ultra-rotating Myers-Perry(MP) black holes can be regarded as
the one similar to the GL instability~\cite{Dias:2010eu}.
The nonlinear dynamical evolution of the GL instability 
has been numerically simulated in Ref.~\cite{Lehner:2010pn} 
but as for the final state it is still under discussion. 
From view points of the membrane paradigm~\cite{Thorne:1986iy} or 
field theory/gravity correspondence~
\cite{Bhattacharyya:2008jc,Emparan:2009cs,Emparan:2009at}, 
dual or analog fluid models of a black string and the GL instability
have been attracting much attention in recent
 years~\cite{Cardoso:2006ks,Cardoso:2006sj,Camps:2010br}.

For the deeper understanding of gravitational instabilities 
and the duality with fluid models, 
it is useful to investigate responses to the variation
or addition of physical parameters such as
the electric charge, the number of dimensions and so on. 
One of the most fundamental and straightforward extension 
is to add rotation. 
In this paper, we consider the $(D+1)$-dim spacetime given by 
\begin{equation}
[D{\rm -dim ~MP~black~ hole}]\times \mathbb R\,, 
\end{equation}
and in particular focus on the $D=4$ case. 
We call these spacetimes {\em MP black strings} for $D>4$ 
and the {\em Kerr black string} for $D=4$. 

Effects of rotation on the GL instability have been discussed 
by several authors~\cite{Cardoso:2006sj,Kleihaus:2007dg,Dias:2010eu}. 
The results of these previous studies suggest that, in general, 
rotation makes the GL instability stronger. 
We briefly summarize the known facts about 
the effects of rotation on the GL instability or 
instabilities on dual/analog fluid models. 
\begin{itemize}

\item{\bf Ultra-spinning MP black strings~\cite{Cardoso:2006sj}}

When we consider the case of $D>5$ and only one rotational axis, 
the angular momentum of the MP black hole is unbounded. 
For a very large angular momentum, 
the metric reduces to~\cite{Emparan:2003sy} 
\begin{equation}
[D{\rm -dim~ MP~ black hole}]\times
 \mathbb R\sim [(D-2){\rm -dim~ Schwraschild}]\times \mathbb R^3. 
\end{equation}
Since this is precisely the geometry considered by 
Gregory and Laflamme~\cite{Gregory:1993vy}, it is unstable. 
Therefore, we can expect that the GL instability persists 
even if rotation is added.

\item{\bf Co-homogeneity-1 MP black strings}

MP black holes with all the angular momenta being equal in magnitude 
have the co-homogeneity-1 geometry (ie, the hypersurface of a constant 
radial coordinate is homogeneous) in odd dimensions.
Therefore, when we consider the geometry 
given by [co-homogeneity-1 MP]$\times \mathbb R$, 
it also has the co-homogeneity-1 symmetry in $D$ dimensions as 
in the case of non-rotating black strings. 
We call these spacetimes co-homogeneity-1 MP black strings.
In Ref.~\cite{Kleihaus:2007dg}, 
from an analysis of stationary perturbations on 
co-homogeneity-1 MP black strings, 
it was predicted that the GL instability persists until 
extremality. 
In Ref.~\cite{Dias:2010eu}, the dispersion relation 
of the instability was numerically calculated and 
the prediction in Ref.~\cite{Kleihaus:2007dg} was confirmed. 
The threshold wavelength and the instability time-scale become shorter 
as the rotation increases, that is, rotation makes the GL instability 
stronger for co-homogeneity-1 MP black strings. 

\item{\bf Rayleigh-Plateau instability\cite{Caldarelli:2008mv}}

The Rayleigh-Plateau instability in a rotating fluid tube 
was discussed in Ref.~\cite{Caldarelli:2008mv} as a system 
dual to an MP black string. 
It is reported that the instability strength and threshold wavenumber 
increase as the rotation increases.

\item{\bf Self gravitating cylinder~\cite{1982ApSS:83:209K}}

We can find a similar destabilizing effect of rotation 
on a self-gravitating cylinder in 
the 4-dimensional Newtonian gravity~\cite{1982ApSS:83:209K}. 

\end{itemize}

In this paper, 
we consider the Kerr black string with $D=4$. 
In this case, 
since Kerr black hole is not co-homogeneity-1, 
it is very difficult to solve 
full perturbation equations.
Thus, we treat the rotation in a perturbative way 
by introducing a small parameter representing the rotation,
that is, we consider only slow rotating black strings. 
We find that the rotation in the present case also
makes the GL instability stronger.

This paper is organized as follows. 
In Sec.~\ref{sec:thermo}, we give a thermo-dynamical 
prediction comparing the entropy 
of a Kerr black string with that of a 5-dim MP black hole. 
Equations for the metric perturbations to second order
in the rotation parameter are derived in Sec.~\ref{sec:pert}
and the results of the numerical calculation are presented 
in Sec.~\ref{sec:result}. Some messy expressions 
are deferred to Appendices~\ref{sec:metom0} and \ref{sec:expomn0}. 
A minimum review of spherical harmonics is given in Appendix~\ref{sec:SH}. 
We set the $D+1$-dimensional gravitational constant and 
the speed of light equal to unity. 

\section{Thermodynamics}
\label{sec:thermo}

First, we discuss the instability of 
stationary black objects from the thermo-dynamical point of view. 
We consider a Kerr black string with the length $L$ along 
the $z$-direction and compare this with the 5-dimensional MP black hole 
which has the same total mass and the angular momentum
 as the Kerr black string. 
Since the horizon area is regarded as the entropy in black hole 
thermodynamics, if the horizon area of the Kerr black string is 
smaller than that of the MP black hole, 
the Kerr black string is thermodynamically less favorable
than the MP black hole. 
Then, it is expected that the Kerr black string becomes
unstable to a perturbation with the wavelength $L$ along the
$z$-direction. 

The metric of a Kerr black string is 
written in the form,
\begin{eqnarray}
ds^2&=&-\frac{\triangle-a^2\sin^2\theta}{\Sigma}dt^2
-2a\sin^2\theta\frac{r^2+a^2-\triangle}{\Sigma}dtd\phi
\nonumber\\
&&+\frac{(r^2+a^2)^2-\triangle a^2\sin^2\theta}{\Sigma}\sin^2\theta
d\phi^2+\frac{\Sigma}{\triangle}dr^2+\Sigma d\theta^2+dz^2, 
\end{eqnarray}
where $0\leq z\leq L$ with $z=0$ and $z=L$ being identified,
and 
\begin{eqnarray}
\Sigma &=&r^2+a^2\cos^2\theta, \\
\triangle &=&r^2-2Mr+a^2=(r-r_+)(r-r_-). 
\end{eqnarray}
The $z=$const., 4-dimensional section is 
given by the metric of a Kerr black hole. 
The tonal mass $E$ and the total angular momentum $J$ 
of the Kerr black string are given by
\begin{equation}
E=LM\,,\quad J=Ea\,. 
\end{equation}
The horizon area is given by 
\begin{equation}
\frac{A_{\rm BS}}{E^{3/2}}
=\frac{4\pi(r_+^2+a^2)L}{E^{3/2}}
=8\pi\frac{1}{l}\left(1+\sqrt{1-j^2l^2}\right), 
\end{equation}
where $l$ and $j$ are non-dimensional quantities defined by 
\begin{equation}
l=\frac{L}{E^{1/2}}\,,
\quad
j=\frac{J}{E^{3/2}}\,. 
\end{equation}

The metric of an MP black hole with single rotation 
can be written as
\begin{eqnarray}
ds^2&=&-dt^2+\frac{M_{\rm MP}}{\Sigma}
\left(dt^2-a_{\rm MP}\sin^2\theta d\phi\right)^2
+\frac{\Sigma}{\triangle_{\rm MP}}dr^2
\nonumber\\
&+&\Sigma d\theta^2+(r^2+a_{\rm MP}^2)\sin^2\theta d\phi^2
+r^2\cos^2\theta dw^2,
\end{eqnarray}
where 
\begin{equation}
\triangle_{\rm MP} =r^2+a_{\rm MP}^2-M_{\rm MP}\,. 
\end{equation}
The total mass and the angular momentum are given by 
\begin{equation}
E=\frac{3}{8}\pi M_{\rm MP}\,,
\quad
J=\frac{2}{3}Ea_{\rm MP}\,. 
\end{equation}
The horizon area is given by 
\begin{equation}
\frac{A_{\rm MP}}{E^{3/2}}
=\frac{2\pi^2M_{\rm MP}r_0}{E^{3/2}}
=\frac{16}{3}\pi \sqrt{\frac{8}{3\pi}-\frac{9}{4}j^2}\,, 
\end{equation}
where we have assumed $M_{\rm MP}>a_{\rm MP}^2$ and 
$r_0$ is the horizon radius given by 
\begin{equation}
r_0=\sqrt{M_{\rm MP}-a_{\rm MP}^2}\,.
\end{equation}

We define the critical length $L_{\rm crit}$ by 
the equality of the horizon area, that is, 
\begin{equation}
A_{\rm BS }=A_{\rm MP}~~{\rm for}~~L=L_{\rm crit}. 
\end{equation}
Then, fixing the total mass $E$, 
we find 
\begin{equation}
\left\{
\begin{array}{rcl}
A_{\rm BS}\geq A_{\rm MP}&{\rm for }&L\leq L_{\rm crit}\\
A_{\rm BS}< A_{\rm MP}&{\rm for }&L> L_{\rm crit}\\
\end{array}
\right., 
\end{equation}
where
\begin{equation}
\frac{L_{\rm crit}}{\sqrt{E}}
=\frac{3\sqrt{3\pi}}{16}\sqrt{32-27j^2\pi}=:l_{\rm crit}. 
\end{equation}
Hence we expect that Kerr black strings are unstable 
against a perturbation with the wavelength $L>L_{\rm crit}$. 
The fact that the value of $l_{\rm crit}$ is a decreasing function of $j$
suggests that the 
critical wavelength of the GL instability becomes shorter, and hence 
the critical wavenumber $\mu_{\rm crit}$ becomes larger 
if we add rotation. 
In the subsequent sections, by calculating the metric perturbations
explicitly, we confirm this observation.

\section{perturbation equations}
\label{sec:pert}
To avoid all possible complexities due to the coordinate singularity on 
the horizon, we use the metric in the Kerr-Schild coordinates given by 
\begin{eqnarray}
ds^2&=&-\frac{\triangle-a^2\sin^2\theta}{\Sigma}dv^2+2dvdr
-\frac{2a\sin^2\theta(r^2+a^2-\triangle)}{\Sigma}dvd\chi
-2a\sin^2\theta d\chi dr
\nonumber\\
&&+\frac{(r^2+a^2)^2-\triangle a^2\sin^2\theta}{\Sigma}
\sin^2\theta d\chi^2+\Sigma d\theta^2+dz^2. 
\end{eqnarray}
In Einstein gravity, the vacuum field equations
are given by 
\begin{equation}
R_{\mu\nu}=0\,. 
\end{equation}
Here, we consider linear metric perturbations on 
the Kerr black string background. 
The metric may be decomposed as 
\begin{equation}
g=g^{(0)}+\hat h +\cdots,  
\end{equation}
where $g^{(0)}$ is the metric of the Kerr black string. 

Imposing the transverse traceless gauge condition,
\begin{equation}
\nabla_\mu \hat h^\mu_{~\nu}=0,~~\hat h^\mu_{~\mu}=0, 
\end{equation}
we can rewrite the metric perturbation equations as
\begin{equation}
\triangle_{\rm L}\hat h^{\mu\nu}
:=\nabla^2\hat h^{\mu\nu}
+2R^{\mu~\nu}_{~\alpha~\beta}\hat h^{\alpha\beta}=0\,, 
\end{equation}
where $\triangle_{\rm L}$ is the so-called Lichnerowicz operator. 

As in the analysis of the original GL instability, 
we assume 
\begin{equation}
\hat h^{\mu z}=0\,,
\quad
\hat h^{zz}=0\,.
\end{equation}
Performing Fourier expansion 
along the symmetric spatial directions, 
we have 
\begin{equation}
\hat h^{ab}
= e^{i(\mu z+m\phi)}e^{\Omega v}h^{ab}(r,\theta), 
\end{equation}
where the subscripts $a$ and $b$ indicate the 
coordinates $(v, r, \theta,\chi)$ and we have assumed 
that the frequency is pure imaginary as in Ref.~\cite{Dias:2010eu}. 
Hereafter we concentrate on the $m=0$ case. 
The assumption of a pure imaginary frequency 
is correct for the non-rotating case~\cite{Gregory:1993vy}.
In our case, the validity of this assumption may be 
supported by the existence of such a solution. 
The equations of motion can be rewritten as 
\begin{equation}
(\triangle_{\rm L} \hat h)^{ab}=-\mu^2 \hat h^{ab}
+(\mathcal L \hat h)^{ab}=0~\Leftrightarrow~
-\mu^2 \tilde h^{ab}+(\mathcal L \tilde h)^{ab}=0, 
\label{eq:eqforh}
\end{equation}
where 
\begin{equation}
\tilde h^{ab}:=e^{\Omega v}h^{ab} 
\end{equation}
and $\mathcal L$ is the Lichnerowicz operator 
in the Kerr spacetime. 
Gauge conditions are given by 
\begin{equation}
\nabla_a \tilde h^{ab}=0\,,
\quad
h^a_{~a}=0\,. 
\label{eq:gaugecond}
\end{equation}
We solve these equations as an eigenvalue problem 
for $\mu$ with a fixed $\Omega$.

As mentioned in Sec.~\ref{sec:intro}, 
we consider the slowly rotating case 
by introducing a small parameter $\epsilon$. 
We specify the deviation from the non-rotating case by
\begin{eqnarray}
M&=&\hat M+\epsilon^2 \hat M, \\
a&=&2\epsilon \hat M. 
\end{eqnarray}
In the limit $\epsilon\rightarrow0$, 
the metric of a $z=$const. hyper-surface becomes 
the Schwarzschild metric. 
Since 
\begin{equation}
\triangle=(r-2\hat M)(r-2\hat M\epsilon^2), 
\end{equation}
we have 
\begin{equation}
r_+=2\hat M,~~r_-=\epsilon^2 r_+. 
\end{equation}
In this way, the coordinate radius of the horizon
is unchanged by adding the rotation. 
This fact makes our analysis much simpler
because the coordinate value at the horizon boundary 
does not depend on the order of $\epsilon$. 

Fixing the value of $\Omega$, 
we expand $\mathcal L$, $\mu^2$ and $h$ as
\begin{eqnarray}
\mathcal L&=&\mathcal L_0+\epsilon \mathcal L_1+\epsilon^2
\mathcal L_2+\cdots\,, \\
\mu^2&=&\mu_0^2+\epsilon\mu_1^2+\epsilon^2\mu_2^2+\cdots\,, \\
\tilde h&=&\tilde h_0+\epsilon \tilde h_1+\epsilon^2 \tilde h_2+\cdots\,. 
\end{eqnarray}
Since the value of $\mu$ should not depend on 
the direction of the rotation, we have $\mu_1^2=0$. 
Thus our goal is to get the value of $\mu_2^2$. 
The equations of motion can be schematically 
described at each order of $\epsilon$ as
\begin{eqnarray}
\epsilon^0&:~&-\mu_0^2\tilde h_0^{ab}+(\mathcal L_0\tilde h_0)^{ab}=0\,,
\label{eq:zeroth} \\
\epsilon^1&:~&-\mu_0^2\tilde h_1^{ab}+(\mathcal L_0\tilde h_1)^{ab}
+(\mathcal L_1\tilde h_0)^{ab}=0\,,
\label{eq:first} \\
\epsilon^2&:~&-\mu_0^2\tilde h_2^{ab}+(\mathcal L_0\tilde h_2)^{ab}
+(\mathcal L_1\tilde h_1)^{ab}+(\mathcal L_2\tilde h_0)^{ab}
-\mu_2^2\tilde h_0^{ab}=0\,. 
\label{eq:secondeq}
\end{eqnarray}

In the following subsections, we find that there exist
a single master variable that satisfies a second-order
ordinary differential equation at each order of $\epsilon$. 
The master variables must be carefully chosen so that the 
equation can be easily solved by numerical integration. 
An inappropriate choice of master variables may cause $0/0$ terms 
in numerical integration as reported in Ref.~\cite{Gregory:1994bj}. 
For this reason, we treat the $\Omega=0$ case 
separately from $\Omega\neq0$ cases. 

\subsection{$\Omega=0$ case}

\subsubsection{Master Equations at order $\epsilon^0$ and $\epsilon^1$}

As the same as the original work by Gregory and Laflamme, 
we consider only $s$-wave perturbations at order $\epsilon^0$. 
The metric perturbation can be written as 
\begin{equation}
h^{ab}_0=
\begin{pmatrix}
h^{vv}_0(r)&h^{vr}_0(r)&0&0\\
h^{vr}_0(r)&h^{rr}_0(r)&0&0\\
0&0&h^{\theta\theta}_0(r)&0\\
0&0&0&\frac{h^{\theta\theta}_0(r)}{\sin^2\theta}
\end{pmatrix}. 
\end{equation}
Here we introduce $H_-$ defined by 
\begin{equation}
H_{-}:=\left(1-\frac{r_+}{r}\right)h^{vv}_0-2h^{vr}_0\,,
\end{equation}
which is found to be a good master variable at order $\epsilon^0$.
Using the gauge conditions (\ref{eq:gaugecond}) and 
the equations of motion (\ref{eq:zeroth}),
we can derive the second-order ordinary differential equation 
for $H_-$\,,
\begin{equation}
H_-''= \left(\frac{3 \mu_0 ^2 r^2+2}{\mu_0 ^2 r^3+2 r-2 r_+}
+\frac{1}{r_+-r}-\frac{4}{r}\right)H_-'+\frac{\mu_0 ^2 r
    \left(\mu_0 ^2 r^3+4 r-8 r_+\right)}{(r-r_+) 
    \left(\mu_0 ^2 r^3+2 r-2 r_+\right)}H_-. 
    \label{eq:om0e0}
\end{equation}
All the components of the metric perturbation can be expressed 
in terms of $H_-$. The explicit expressions are
given in Appendix~\ref{sec:metom0e0}.

The metric perturbation at order $\epsilon^1$ 
is induced by the perturbation at order $\epsilon^0$ 
as given by Eq.~(\ref{eq:first}).
The only nontrivial induced components are found to be
$h^{v\chi}_1$ and $h^{r\chi}_1$:
\begin{equation}
h^{ab}_1=
\begin{pmatrix}
0&0&0&h^{v\chi}_1(r)\\
0&0&0&h^{r\chi}_1(r)\\
0&0&0&0\\
h^{v\chi}_1(r)&h^{r\chi}_1(r)&0&0
\end{pmatrix}. 
\end{equation}
Defining $f(r):=h^{v\chi}_1(r)$ as a master variable, 
we can derive the following equation:
\begin{eqnarray}
&&f''=\left(\frac{2}{r_+-r}-\frac{2}{r}\right) f'+\frac{ \left(\mu_0
   ^2 r^3-2 r_+\right)}{r^2 (r-r_+)}f+S_1\,;
   \\
&&~S_1=
\frac{r_+  \left(3 r r_+^3 \left(\mu_0 ^2 r^2+12\right)
-4 r^2 r_+^2 \left(\mu_0 ^2 r^2+7\right)+r^3
   r_+ \left(\mu_0 ^2 r^2-12\right)
+12 r^4-12 r_+^4\right)}{2 r^3 (r-r_+)^3 \left(\mu_0 ^2 r^3+2 r-2
   r_+\right)}H_-'
   \nonumber\\
&&~+
\frac{r_+  \left(r_+^3 \left(10-3 \mu_0 ^2 r^2\right)
+2 r r_+^2 \left(\mu_0 ^2 r^2-11\right)
+r^2 r_+ \left(7 \mu_0 ^2 r^2+6\right)
-2 \left(2 \mu_0 ^2 r^5+r^3\right)\right)}{r^3 (r-r_+)^3 
\left(\mu_0^2 r^3+2 r-2 r_+\right)}H_-\,.
\nonumber
\end{eqnarray}

\subsubsection{Master equation at order $\epsilon^2$}
\label{sec:eq2l0}

Equations at order $\epsilon^2$ are much more complicated 
than those to order $\epsilon^1$.
However, as we will see soon, we do not need to fully 
solve the perturbation equations to get the value of $\mu_2$. 
Instead it is enough to focus on $\ell=0$ modes 
in the spherical harmonics expansion. 
A minimal review of tensor and vector spherical harmonics 
is given in Appendix~\ref{sec:SH}. 
In Eq.~(\ref{eq:secondeq}), the first and second terms correspond
to the homogeneous part of the coupled differential equations.
The operator of this part is the same as
that at order $\epsilon^0$. This implies that
different spherical harmonic modes of $h_2$ do not couple with each other
within this homogeneous part. 
In addition, the final term proportional to $\mu_2^2$
has only the $\ell=0$ mode. 
Therefore, the value of $\mu_2^2$ can be obtained
by solving only the $\ell=0$ mode of Eq.~(\ref{eq:secondeq}). 

Extracting the $\ell=0$ mode from Eq.~(\ref{eq:secondeq}),
we obtain 
\begin{equation}
-\mu_0^2h_{2,0}^{ab}+(\mathcal L_0h_{2,0})^{ab}
+\left[(\mathcal L_1h_1)^{ab}
+(\mathcal L_2h_0)^{ab}\right]_{\ell=0}-\mu_2^2h_0^{ab}=0\,, 
\end{equation}
where
\begin{equation}
h_{2,0}^{ab}(r):=\left[h_2^{ab}(r)\right]_{\ell=0}\,,
\end{equation}
which can be expressed as
\begin{equation}
h^{ab}_{2,0}=
\begin{pmatrix}
h^{vv}_2(r)&h^{vr}_2(r)&0&0\\
h^{vr}_2(r)&h^{rr}_2(r)&0&0\\
0&0&h^{\theta\theta}_2(r)&0\\
0&0&0&\frac{h^{\theta\theta}_2(r)}{\sin^2\theta}
\end{pmatrix}. 
\end{equation}
We also extract the $\ell=0$ mode from the gauge conditions,
\begin{equation}
\left[\nabla_a h^a_{~b}\right]_{l=0}=0~,~~(h_{2,0})^a_{~a}=0\,.
\end{equation}

Here in analogy with the case of order $\epsilon^0$,
we introduce the following master variable:
\begin{equation}
H_{2-}:=\left(1-\frac{r_+}{r}\right)h^{vv}_2-2h^{vr}_2\,. 
\end{equation}
Then we obtain the master equation,
\begin{eqnarray}
&&H_{2-}''=
 \left(\frac{3 \mu_0 ^2 r^2+2}{\mu_0 ^2 r^3+2 r-2 r_+}
 +\frac{1}{r_+-r}-\frac{4}{r}\right)H_{2-}'
\nonumber\\
&&\hspace{40mm}
   +\frac{ \mu_0 ^2 r \left(\mu_0 ^2 r^3+4 r-8
   r_+\right)}{(r-r_+) \left(\mu_0 ^2 r^3+2 r-2 r_+\right)}
   H_{2-}+S_2\,;
\label{eq:om0e2}
\\
&&~S_2=
   \frac{4  \mu_0 ^2 r r_+^2}{\mu_0 ^2 r^3+2 r-2 r_+}f'
   +\frac{4  \mu_0 ^2 r_+^2 (2
   r-r_+)}{(r-r_+) \left(\mu_0 ^2 r^3+2 r-2 r_+\right)}f
   \nonumber\\
   &&
\quad+\frac{1}{3 r^4 (r-r_+)^2 (\mu_0 ^2 r^3+2 r-2 r_+)^2}
   \Biggl(r_+ \Bigl[-3 \mu_0 ^4 r^8(r-r_+) (3 r+r_+)
   \nonumber\\
   &&
\quad+\mu_0 ^2 r^3 (18 r^5+4 r^4 r_+-73 r^3 r_+^2+14 r^2 r_+^3+31 r
   r_+^4-12 r_+^5)
   \nonumber\\
   &&
\quad+4 r_+ (r-r_+) (19 r^4-33 r^3 r_+-3 r^2 r_+^2+17 r r_+^3-6
   r_+^4)\Bigr]+6 \mu_2^2 r^6 (2 r-3 r_+) (r-r_+)^2\Biggr)H_-'
   \nonumber\\
   &&
\quad+\frac{1}{3 r^4(r-r_+)^2 (\mu_0 ^2 r^3+2 r-2 r_+)^2}
    \Biggl(3 \mu_2^2 r^5 (r-r_+) 
    \Bigl[-4 r r_+ (\mu_0 ^2r^2+6)
   \nonumber\\
   &&
\quad+r^2 (\mu_0 ^4 r^4+4 \mu_0 ^2 r^2+8)+16 r_+^2\Bigr]
   +r_+ \Bigl[2 r_+^5 (3 \mu_0 ^2
   r^2-44)+r r_+^4 (-3 \mu_0 ^4 r^4+44 \mu_0 ^2 r^2+208)
   \nonumber\\
   &&
\quad-2 r^2 r_+^3 (2 \mu_0 ^4 r^4+35 \mu_0 ^2
   r^2-24)+r^4 r_+ (-2 \mu_0 ^6 r^6+23 \mu_0 ^4 r^4+84 \mu_0 ^2 r^2+200)
   \nonumber\\
   &&
\quad+3 r^5 (\mu_0 ^6 r^6-8 \mu_0 ^4 r^4-16
   \mu_0 ^2 r^2-16)
   -r^3 r_+^2 (\mu_0 ^6 r^6-26 \mu_0 ^4 r^4+88 \mu_0 ^2 r^2+320)
   \Bigr]\Biggr)H_-\,. 
   \nonumber
   \end{eqnarray}
Explicit expressions for the metric components in terms of 
$H_{2-}$ are given in Appendix~\ref{sec:metom0e2}. 
   
\subsubsection{Asymptotic behavior}
\label{sec:infinity}

In order to solve the master equations, 
we need to specify the boundary conditions. 
In the limit $r\rightarrow \infty$, 
Eq.~(\ref{eq:om0e0}) can be approximated as 
\begin{equation}
H_-''+\frac{2}{r}H_-'-\mu_0^2H_-=0\,. 
\end{equation}
Hence the general solution at $r\to\infty$ is given by 
\begin{equation}
H_-=C_1\frac{e^{-\mu_0 r}}{r}+C_2\frac{e^{\mu_0 r}}{r}\,, 
\end{equation}
where we may assume $\mu_0>0$ without loss of generality.
Then the regularity requires $C_2=0$.
Thus $H_-$ decays exponentially at infinity,
and so do all the components of the metric at order $\epsilon^0$.
At order $\epsilon$, the homogeneous solution of the master equation
is found to have the similar asymptotic behavior.
Since the source term is given in terms of $H_-$ and $H_{-}'$,
this implies that the master variable $f(r)$, and hence
all the components of the metric perturbation at order $\epsilon$ 
also decay exponentially at infinity.
The same is true for the master equation at order $\epsilon^2$
since it has the same form as the master equation at order $\epsilon^0$.
Thus all the components of the metric perturbations to order
$\epsilon^2$ must decay exponentially at the infinity.

\subsubsection{Near the horizon}
\label{sec:nearho}

From the regularity at the horizon, 
we find $H_-=\mathcal O((r-r_+)^0)$, $f=\mathcal O((r-r_+)^0)$
and $H_{2-}=\mathcal O((r-r_+)^0)$. 
Expanding the zeroth and first order master equations 
near the horizon, we obtain the following 
asymptotic solutions for $H_-$ and $f$:
\begin{eqnarray}
H_-&=&\alpha\left(1+
\left(\mu_0 ^2 r_+-\frac{4}{r_+}\right) (r-r_+)
+\left(\frac{\mu_0 ^4 r_+^2}{4}-2 \mu_0
   ^2+\frac{8}{r_+^2}\right) (r-r_+)^2\right)
   \nonumber\\
   &&
   +\mathcal O((r-r_+)^3), 
\\
f&=&\beta -\frac{(r-r_+)
\left(-6 \beta  \mu_0 ^2 r_+^3+12 \beta  r_++\mu_0 ^4 r_+^4-19 \mu_0 ^2
   r_+^2+94\right)}{12 r_+^2}+\mathcal O\left((r-r_+)^2\right), 
\end{eqnarray}
where $\alpha$ and $\beta$ are constants to be determined. 
Without loss of generality, we may set $\alpha=1$ because 
Eq.~\eqref{eq:om0e0} is a homogeneous linear equation.
The eigenvalue $\mu_0^2$ is determined so that $H_-$ decays 
exponentially at infinity, and the constant $\beta$
is determined so that $f$ decays exponentially at infinity. 

Since the zeroth order equation~\eqref{eq:om0e0} and 
the second order equation~\eqref{eq:om0e2} have the same
form except for the presence of the source term $S_2$ for
the latter, one can add the zeroth order solution $H_-$
to any particular solution $H_{2-}$ of Eq.~\eqref{eq:om0e2}. 
Using this freedom, we can set $H_{2-}(r_+)=0$ by
adding a term proportional to $H_-$ appropriately.
Then expanding Eq.~\eqref{eq:om0e2} near the horizon,
we find
\begin{eqnarray}
H_{2-}&=&(r-r_+) \left(4 \beta +r_+ \left(\frac{\mu_0 ^2}{3}
+\mu_2^2\right)-\frac{7}{r_+}\right)+\mathcal O\left((r-r_+)^2\right). 
\end{eqnarray}
As before, the value of $\mu_2^2$ is determined so that 
$H_{2-}$ exponentially decays at infinity. 

\subsection{$\Omega\neq0$ case}
\label{sec:nonstationary}

In the $\Omega\neq0$ case if we would use the same master variables 
as those in the $\Omega=0$ case, we would suffer from $0/0$ terms 
in the master equations at order $\epsilon^0$ and $\epsilon^2$,
as pointed out in Ref.~\cite{Gregory:1994bj}.
To avoid numerical errors due to these $0/0$ terms, 
as in Ref.~\cite{Gregory:1994bj}, 
we rewrite each master equation at order $\epsilon^0$ and 
$\epsilon^2$ as two coupled first-order differential equations 
by introducing an additional variable. 

At order $\epsilon^0$, we define a function $H$ as 
\begin{eqnarray}
H&:=&-h_0^{vr}+\frac{1}{1-\frac{r_+}{r}}h_0^{rr}. 
\end{eqnarray}
Then, we obtain the following equations for $H_-$ and $H$: 
\begin{eqnarray}
H_-'&=&\frac{\left(\mu_0 ^4 r^3+2 \mu_0 ^2 (r-r_+)
-4 \Omega ^2 r\right)}{\Omega  \left(\mu_0 ^2
   r^3+r_+\right)}H \nonumber\\
   &+&\frac{\left(\mu_0 ^2 r^3 (3 r_+-2 r (\Omega  r+1))+4 \Omega ^2 r^4-2 r r_+
   (\Omega  r+3)+6 r_+^2\right)}{2 r (r-r_+) \left(\mu_0 ^2 r^3+r_+\right)}H_-\,, 
   \label{eq:zeroomn01}
   \\
H'&=&
\frac{\Omega  r  \left(4 \mu_0 ^2 r^3 (r-r_+)
+4 \Omega ^2 r^4-r_+^2\right)}{4 (r-r_+)^2 \left(\mu_0 ^2
   r^3+r_+\right)}H_-\nonumber\\
   &-&\frac{ \left(\mu_0 ^2 r^3 (2 r (\Omega  r+1)+r_+)
+4 \Omega ^2 r^4+2 r r_+ (\Omega  r+2)-2
   r_+^2\right)}{2 r (r-r_+) \left(\mu_0 ^2 r^3+r_+\right)}H\,. 
   \label{eq:zeroomn02}
   \end{eqnarray}
These equations are free from $0/0$ terms.
The components of the metric perturbation 
are expressed in terms of $H_-$ and $H$ as shown in 
Appendix~\ref{sec:expomn0}. 

At order $\epsilon^1$, we can use the same master variable as
the one in the $\Omega=0$ case, $f:=h^{v\chi}_1$.
It satisfies
\begin{eqnarray}
f''&=&\left(\frac{\mu_0 ^2}{\mu_0 ^2 r+4 \Omega}+\frac{2 (\Omega
   r+1)}{r_+-r}-\frac{3}{r}\right)f' 
   +\frac{\mu_0 ^4 r^4-2 \mu_0 ^2 r r_+-24 \Omega^2 r^2-12
   \Omega r_+}{r^2 (r-r_+) \left(\mu_0 ^2 r+4 \Omega\right)}f 
   \nonumber\\
   &&
   +
   A_H(r;r_+,\mu_0,\Omega)H
   +A_{H_-}(r;r_+,\mu_0,\Omega)H_-\,, 
   \label{eq:firstomn0}
   \end{eqnarray}
where explicit expressions of
 $A_H(r;r_+,\mu_0,\Omega)$ and $A_{H_-}(r;r_+,\mu_0,\Omega)$
are given in Appendix~\ref{sec:expomn0}. 
The other component $h^{r\chi}_1$ of the 
metric perturbation is also expressed in terms of $f$
in Appendix~\ref{sec:expomn0}. 

At order $\epsilon^2$, we define $H_2$ as 
\begin{eqnarray}
H_2&:=&-h_2^{vr}+\frac{1}{1-\frac{r_+}{r}}h_2^{rr}. 
\end{eqnarray}
Then, we can derive 
   \begin{eqnarray}
   H_{2-}'&=&
   \frac{ (4 \Omega ^2 r^4+\mu_0 ^2 (3 r_+-2 r (\Omega  r+1)) r^3-2 (\Omega  r+3)
   r_+ r+6 r_+^2)}{2 r (r-r_+) (\mu_0 ^2 r^3+r_+)}
   H_{2-}
   \nonumber\\
   &&
   +\frac{ (r^3 \mu_0 ^4+2 (r-r_+)
   \mu_0 ^2-4 \Omega ^2 r)}{\Omega  (\mu_0 ^2 r^3+r_+)}H_2
   \nonumber\\
   &&
   +\frac{4  (r \mu_0 ^2+2 \Omega )
 (r-r_+) r_+^2}{r (r \mu_0 ^2+4 \Omega ) (\mu_0 ^2 r^3+r_+)}
   f'
   +\frac{ (8 r (r \mu_0 ^2+\Omega  (\mu_0 ^2 r^2+2 \Omega  r+4))
   r_+^2-4 (r \mu_0 ^2+6 \Omega ) r_+^3)}{r^2 (r \mu_0 ^2+4 \Omega ) (\mu_0 ^2
   r^3+r_+)}
   f
   \nonumber\\
   &&
   +B_H(r;r_+,\mu_0,\mu_2,\Omega)H
   +B_{H_-}(r;r_+,\mu_0,\mu_2,\Omega)H_-, 
   \label{eq:secondomn01}
   \end{eqnarray}
   \begin{eqnarray}
   H_2'&=&
   \frac{ \Omega  r (4 \Omega ^2 r^4+4 \mu_0 ^2
   (r-r_+) r^3-r_+^2)}{4 (r-r_+)^2 (\mu_0 ^2 r^3+r_+)}
   H_{2-}
   \nonumber\\
   &&
   -\frac{ (4 \Omega ^2
   r^4+\mu_0 ^2 (2 r (\Omega  r+1)+r_+) r^3
+2 (\Omega  r+2) r_+ r-2 r_+^2)}{2 r (r-r_+) (\mu_0 ^2 r^3+r_+)}
   H_2
   \nonumber\\
   &&
   +\frac{2  \Omega  (2 (r \mu_0 ^2+\Omega ) r^2+r_+) r_+^2}
{r (r \mu_0 ^2+4 \Omega) (\mu_0 ^2 r^3+r_+)}
   f'
   +\frac{2  \Omega  (2 (r \mu_0 ^2+2 \Omega  (r (r \mu_0
   ^2+\Omega )+2)) r^3-4 \Omega  r_+ r^2+r_+^2) r_+^2}
{r^2 (r \mu_0 ^2+4 \Omega) (r-r_+) (\mu_0 ^2 r^3+r_+)}
   f
   \nonumber\\
   &&
   +C_H(r;r_+,\mu_0,\mu_2,\Omega)H
   +C_{H_-}(r;r_+,\mu_0,\mu_2,\Omega)H_-\,, 
   \label{eq:secondomn02}
   \end{eqnarray}
where explicit expressions of
$B_H(r;r_+,\mu_0,\mu_2,\Omega)$, $B_{H_-}(r;r_+,\mu_0,\mu_2,\Omega)$, 
$C_H(r;r_+,\mu_0,\mu_2,\Omega)$ and $C_H(r;r_+,\mu_0,\mu_2,\Omega)$
are given in Appendix~\ref{sec:expomn0}. 
The components of the metric perturbation 
are also shown in Appendix~\ref{sec:expomn0}. 

At infinity, all these variables must exponentially decay. 
We can derive asymptotic solutions near the horizon 
as in the $\Omega=0$ case. 
We describe these asymptotic solutions near the horizon 
in Appendix~\ref{sec:expomn0}.

\section{results and discussion}
\label{sec:result}
Since a second-order ordinary differential equation 
can be regarded as a set of two coupled 
first-order differential equations, 
we have six coupled first-order ordinary differential equations
for each value of $\Omega$. 
We have to determine the values of $\mu_0$, $\beta$ and
$\mu_2$ so that the metric perturbation is regular
at horizon and decays exponentially at infinity. 
It can be done numerically by a shooting method. 
We have solved the ordinary differential equations 
from the horizon to infinity 
using the 4-th order Runge-Kutta method and 
searched for appropriate values of $\mu_0$, $\beta$ and 
$\mu_2$ for each value of $\Omega$. 

\begin{figure}[h]
\begin{center}
\includegraphics[scale=2]{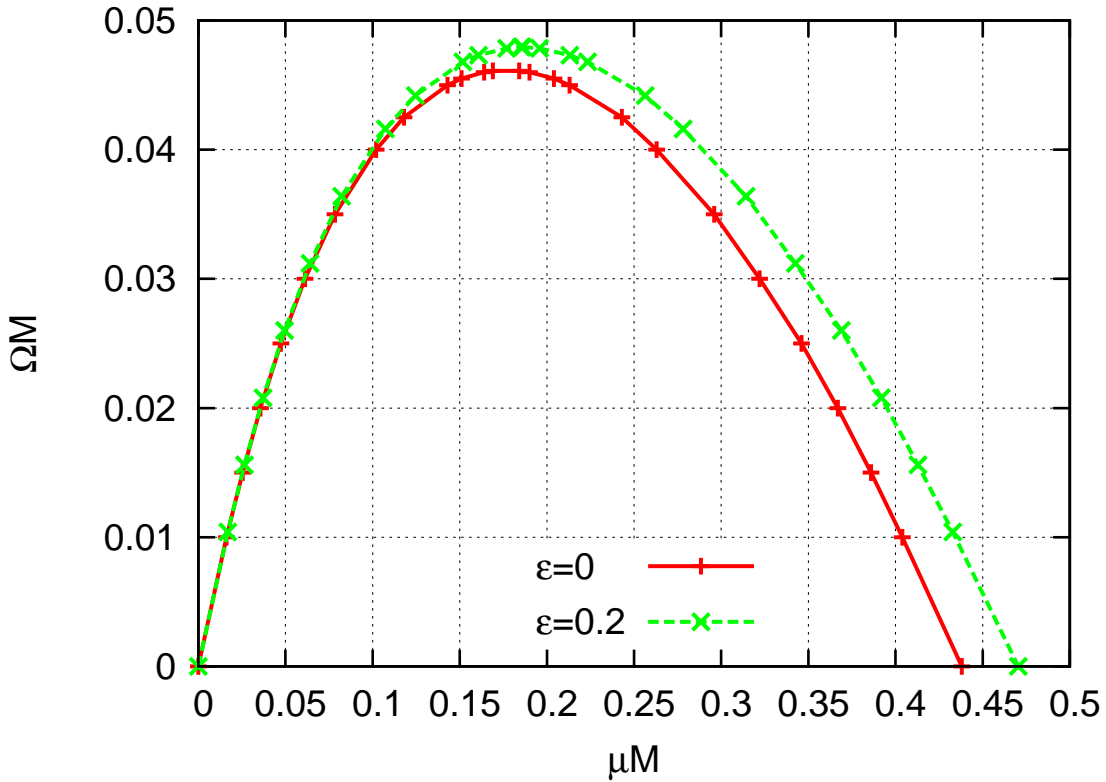}
\caption{The growth rate of the GL instability $\Omega$
as a function of the eigenvalue $\mu(=\sqrt{\mu_0^2+\epsilon^2\mu_2^2})$.
Plotted are the case of $\epsilon=0$(no rotation) and $\epsilon=0.2$.
}
\label{fig:second}
\end{center}
\end{figure}

Here, we show the numerical results by fixing
 the mass parameter $M$ of the Kerr black hole at the 
4-dimensional section, although 
the numerical integration is done by 
fixing the value of $r_+$. 
That is, we plot $\Omega M$ as a function of
$\mu M$, where $\mu=\sqrt{\mu_0^2+\mu_2^2\epsilon^2}$,
in Fig.~\ref{fig:second}, where
\begin{eqnarray}
\Omega M&=&\Omega \hat M(1+\epsilon^2)\,, \\
\mu M&=&\sqrt{\mu_0^2+\mu_2^2\epsilon^2} \hat M(1+\epsilon^2)
=\mu_0\hat M\left(1+\left(1+\frac{1}{2}
\frac{\mu_2^2}{\mu_0^2}\right)\epsilon^2\right)
+\mathcal O(\epsilon^3)\,. 
\end{eqnarray}

It is clear from the figure that both the critical wavelength
 and the growth time-scale of the instability become shorter as
the rotation is increased. 
In other words, rotation makes 
the black string system more unstable. 
This result is consistent with previous works in similar
situations as listed in Sec.~\ref{sec:intro}. 

Although our results are consistent with 
all the previous results, an intuitive understanding of our results
seems rather unclear. 
If we fix the mass parameter of the Kerr black hole 
in the 4-dimensional section, the horizon area 
decreases as the angular momentum is increased. 
In this sense, the typical time-scale or length-scale becomes shorter.
This is consistent with our and previous results,
and this was in fact what we obtained from thermodynamic arguments. 
However, in contrast, if we focus on the surface gravity,
it decreases as the angular momentum is increased. 
In this respect, the typical time-scale or length-scale 
may become longer. Actually, when we add a charge to 
a neutral black brane solution, it is known that
the GL instability gets weaker~\cite{Gregory:1994bj}. 
This is totally opposite to the behavior we obtained.
Apparently we need a better physical picture
of the GL instability with rotation.

\section*{Acknowledgements}
We thank T. Tanaka, G. Kang for helpful discussions and comments.
We made much progress during the APCTP Joint Focus Program: 
Frontiers of Black Hole Physics (December 2010).
We would like to thank all the participants of this workshop 
for fruitful discussions. 
C.Y. is supported by JSPS Grant-in-Aid for  Creative
Scientific Research No.~19GS0219 and JSPS Grant-in-Aid for Scientific
Research (C)  No.~21540276. 
This work was also supported in part by 
Korea Institute for Advanced Study under the KIAS Scholar program,  
by the Grant-in-Aid for the Global COE Program 
``The Next Generation of Physics, Spun from Universality and Emergence''
from the Ministry of Education, Culture, 
Sports, Science and Technology (MEXT) of Japan, 
by JSPS Grant-in-Aid for Scientific Research (A) No.~21244033,
and by JSPS Grant-in-Aid for Creative Scientific Research No.~19GS0219.

\appendix

\section{Spherical Harmonics}
\label{sec:SH}

Full expressions for the tensor spherical harmonics 
can be found in Refs.~\cite{Regge:1957td,1970JMP....11.2203Z,Zerilli:1971wd}. 
Here, we give a minimum review necessary for the present work. 
We represent tensors by using the components of 
the coordinate system $(v,r,\theta,\phi)$, 
where $v$ is an advanced null coordinate. 
First, we introduce the vector spherical harmonics. 
We decompose any covariant vector field $\bm v$ as 
\begin{equation}
\bm v=\sum_{l,m}\left[
A_{lm} \bm a_{lm}
+B_{lm} \bm b_{lm}
+C_{lm} \bm c_{lm}
+D_{lm} \bm d_{lm}
\right], 
\end{equation}
where $A$, $B$, $C$, $D$ are the coefficients for bases
$\bm{a,~b,~c,~d}$ given by
\begin{eqnarray}
\bm a_{lm} &=&-i(Y_{lm},-Y_{lm},0,0)\,,\\
\bm b_{lm} &=&(0,Y_{lm},0,0)\,,\\
\bm c_{lm} &=&\frac{1}{\sqrt{l(l+1)}}
(0,0,\partial_\theta Y_{lm},\partial_\phi Y_{lm})\,,\\
\bm d_{lm} &=&\frac{1}{\sqrt{l(l+1)}}
(0,0,\frac{1}{\sin\theta}\partial_\phi Y_{lm},
-\sin\theta\partial_\theta Y_{lm})\,.
\end{eqnarray}
These bases are orthonormal in the inner product
\begin{equation}
(\bm v, \bm u):=\int\eta^{ab}v_a u_b \dd\Omega, 
\end{equation}
where $\eta$ is the Minkowski metric. 

We focus on the $l=m=0$ mode. 
In this case, we have the following two non-zero bases:
\begin{equation}
\bm a:=\bm a_{00}=\frac{-i}{2\sqrt{\pi}}
(1,~-1,~0,~0)~,~~
\bm b:=\bm b_{00}=\frac{1}{2\sqrt{\pi}}
(0,~1,~0,~0). 
\end{equation}
Thus, we can extract the $l=m=0$ mode 
from a covariant vector $\bm v$ as 
\begin{equation}
[\bm v]_{l=0}=(\bm a, \bm v)\bm a+(\bm b, \bm v)\bm b. 
\end{equation}

In the same way as in the vector spherical harmonics, 
we can decompose a rank-2 covariant tensor $\mathbf v$
as follows:
\begin{eqnarray}
v&=&\sum_{l, m}\Bigl[
A_{lm}\mathbf a_{lm}
+B_{lm}\mathbf b_{lm}
+C_{lm}\mathbf c_{lm}
+D_{lm}\mathbf d_{lm}
+E_{lm}\mathbf e_{lm}
+F_{lm}\mathbf f_{lm}
\nonumber\\
&&
+G_{lm}\mathbf g_{lm}
+H_{lm}\mathbf h_{lm}
+I_{lm}\mathbf i_{lm}
+J_{lm}\mathbf j_{lm}
\Bigr], 
\end{eqnarray}
where $\mathbf{a,~b,~c,~d,~e,~f,~g,~h,~i,~j}$
are bases of the tensor spherical harmonics and 
$A,~B,~C,\cdots,~I,~J$ are the coefficients for these bases. 
Explicit forms of the bases are 
given in Ref.~\cite{Zerilli:1971wd}. 
These bases are orthogonal in the inner product
\begin{equation}
<\mathbf v,~\mathbf u>:=\int\eta^{ac}\eta^{bd}v_{ab}u_{cd}\dd\Omega. 
\end{equation}
For $m=l=0$, 
we have the following four non-zero bases:
\begin{eqnarray}
\mathbf a:=\mathbf a_{0,0}&=&\frac{1}{2\sqrt{\pi}}
\begin{pmatrix}
~1~&-1~&~0~&~0~\\
-1~&~1~&~0~&~0~\\
~0~&~0~&~0~&~0~\\
~0~&~0~&~0~&~0~
\end{pmatrix}, 
\\
\mathbf b:=\mathbf b_{0,0}&=&\frac{i}{2\sqrt{2\pi}}
\begin{pmatrix}
~0~&-1~&~0~&~0~\\
-1~&~2~&~0~&~0~\\
~0~&~0~&~0~&~0~\\
~0~&~0~&~0~&~0~
\end{pmatrix}, 
\\
\mathbf c:=\mathbf c_{0,0}&=&\frac{1}{2\sqrt{\pi}}
\begin{pmatrix}
~0~&~0~&~0~&~0~\\
~0~&~1~&~0~&~0~\\
~0~&~0~&~0~&~0~\\
~0~&~0~&~0~&~0~
\end{pmatrix}, 
\\
\mathbf d:=\mathbf d_{0,0}&=&\frac{r^2}{2\sqrt{2\pi}}
\begin{pmatrix}
~0~&~0~&~0~&~0~\\
~0~&~0~&~0~&~0~\\
~0~&~0~&~1~&~0~\\
~0~&~0~&~0~&\sin^2\theta
\end{pmatrix}. 
\end{eqnarray}
We can extract the $l=m=0$ mode from $\mathbf v$ as
\begin{equation}
[\mathbf v]_{l=0}=<\mathbf a, \mathbf v>\mathbf a+<\mathbf b, \mathbf v>\mathbf b
+<\mathbf c, \mathbf v>\mathbf c+<\mathbf d, \mathbf v>\mathbf d. 
\end{equation}

\section{metric perturbation for $\Omega=0$}
\label{sec:metom0}
Here we give explicit expressions for the components of
the metric perturbation at each order of $\epsilon$
in the case $\Omega=0$.
For notational simplicity, $\epsilon$ is set to unity
in the following expressions.

\subsection{Metric perturbation at order $\epsilon^0$}
\label{sec:metom0e0}
\begin{eqnarray}
h_0^{vv}&=&
\frac{r^2 (2 r-3 r_+) }{(r-r_+) \left(\mu_0 ^2 r^3+2 r-2 r_+\right)}H_-'
+\frac{r\left(\mu_0^2 r^3+4 r-8 r_+\right)}
{(r-r_+) \left(\mu_0 ^2 r^3+2 r-2 r_+\right)} H_- \,,
 \\
h_0^{vr}&=&
\frac{r (2 r-3 r_+) }{2 \mu_0 ^2 r^3+4 r-4 r_+}H_-'
+\frac{ r-3 r_+}{\mu_0 ^2 r^3+2 r-2 r_+}H_-\,, 
\\
h_0^{rr}&=&
\frac{(2 r-3 r_+) (r-r_+)}{2 \mu_0 ^2 r^3+4 r-4 r_+} H_-'
+\frac{(r-3 r_+)   (r-r_+)}{r \left(\mu_0 ^2 r^3+2 r-2 r_+\right)}H_- \,,
\\
h_0^{\theta\theta}&=&\frac{1}{2r^2}H_-\,. 
\end{eqnarray}

\subsection{Metric perturbation at order $\epsilon^1$}
\label{sec:metom0e1}
 \begin{equation}
h_1^{r\chi}=
\frac{r_+(2 r -3 r_+) }{2 \mu_0 ^2 r^4+4 r^2-4 r r_+}H_-'
+\frac{r_+  (r-3r_+)}{r^2 \left(\mu_0 ^2 r^3+2 r-2 r_+\right)}H_-. 
\end{equation}

\subsection{Metric perturbation at order $\epsilon^2$}
\label{sec:metom0e2}
   \begin{eqnarray}
   h^{rr}_2&=&
\frac{(2 r-3 r_+) (r-r_+) }{2\mu_0 ^2 r^3+4 r-4 r_+}H_{2-}'
+\frac{(r-3 r_+) (r-r_+)}{r \left(\mu_0 ^2 r^3+2 r-2 r_+\right)}H_{2-}
   \nonumber\\
   &&
 +\frac{2  r_+^2 (r-r_+)^2}{r \left(\mu_0 ^2 r^3+2 r-2 r_+\right)}{h_1^{v\chi}}'
 +\frac{2
    r_+^2 (r-r_+) (2 r-r_+)}{r^2 \left(\mu_0 ^2 r^3+2 r-2 r_+\right)}h_1^{v\chi}
   \nonumber\\
   &&
 +\Biggl(r_+ \Bigl(-21 \mu_0 ^2 r^7+31 \mu_0 ^2 r^6 r_++2 r^5 (\mu_0 ^2 r_+^2-3)+
   \nonumber\\
   &&
   r^4 r_+ (40-9 \mu_0 ^2
   r_+^2)-r^3 r_+^2 (3 \mu_0 ^2 r_+^2+59)+4 r^2 r_+^3+9 r r_+^4+6
   r_+^5\Bigr)
   \nonumber\\
   &&
   -3 \mu_2^2 r^6 (2 r-3 r_+) (r-r_+)\Biggr)
\frac{H_-'}{6 r^3 (\mu_0 ^2 r^3+2 r-2 r_+)^2}
   \nonumber\\
   &&
   +\Biggl(r_+ \Bigl(-36 \mu_0 ^2 r^7+79 \mu_0 ^2 r^6 r_+-4 r^5 (4 \mu_0 ^2
   r_+^2+9)
   \nonumber\\
   &&
   +8 r^4 r_+ (14-3 \mu_0 ^2 r_+^2)+r^3 r_+^2 (3 \mu_0 ^2
   r_+^2-176)+34 r^2 r_+^3+48 r r_+^4-6 r_+^5\Bigr)
   \nonumber\\
   &&
   -6 \mu_2^2 r^6 (r-3 r_+)
   (r-r_+)\Biggr)\frac{H_-}{6 r^4 (\mu_0 ^2 r^3+2 r-2 r_+)^2}    , 
   \end{eqnarray}
   \begin{eqnarray}
   h^{vv}_2&=&\frac{r^2 (2 r-3 r_+)
   }{(r-r_+) \left(\mu_0 ^2 r^3+2 r-2 r_+\right)}H_{2-}'
   +\frac{ r \left(\mu_0 ^2 r^3+4 r-8
   r_+\right)}{(r-r_+) \left(\mu_0 ^2 r^3+2 r-2 r_+\right)}
   H_{2-}
   \nonumber\\
   &&
   +
   \frac{4  r r_+^2}{\mu_0 ^2 r^3+2 r-2 r_+}
   {h_1^{v\chi}}'
   +\frac{4  r_+^2 (2
   r-r_+)}{(r-r_+) \left(\mu_0 ^2 r^3+2 r-2 r_+\right)}h_1^{v\chi}
   \nonumber\\
   &&
   +\Biggl(r_+ \Bigl(-15 \mu_0 ^2 r^7-3 r r_+^4 (\mu_0 ^2 r^2+3)
   \nonumber\\
   &&
   +2 r^4 r_+ (11 \mu_0 ^2 r^2+5)-r^3 r_+^2 (4 \mu_0 ^2
   r^2+53)+6 r^5+34 r^2 r_+^3+6 r_+^5\Bigr)
\nonumber\\
  && -3 \mu_2^2 r^6 (2 r-3 r_+) (r-r_+)\Biggr)
\frac{H_-'}{3 r(r-r_+)^2 (\mu_0 ^2 r^3+2 r-2 r_+)^2}
   \nonumber\\
   &&
   +\Biggl(r_+ \Bigl(3 r r_+^4 (\mu_0 ^2
   r^2+4)+2 r^2 r_+^3 (41-3 \mu_0 ^2 r^2)
   \nonumber\\
   &&
   -6 r^5 (5 \mu_0 ^2 r^2+4)+r^4 r_+ (61 \mu_0 ^2
   r^2+64)-2 r^3 r_+^2 (11 \mu_0 ^2 r^2+76)-6 r_+^5\Bigr)
   \nonumber\\
   &&
   -6 \mu_2^2 r^6 (r-3 r_+)(r-r_+)\Biggr)
\frac{H_-}{3 r^2 (r-r_+)^2 (\mu_0 ^2 r^3+2 r-2 r_+)^2} , 
   \end{eqnarray}
   \begin{eqnarray}
   h^{vr}_2&=&
   \frac{r (2 r-3 r_+)
   }{2 \mu_0 ^2 r^3+4 r-4 r_+}H_{2-}'
+\frac{ (r-3 r_+)}{\mu_0 ^2 r^3+2 r-2 r_+}
   H_{2-}
  \nonumber\\
  &&  + \frac{2  r_+^2 (r-r_+)}{\mu_0 ^2 r^3+2 r-2 r_+}{h_1^{v\chi}}'
   +\frac{ 2r_+^2(2 r-r_+)}{\mu_0 ^2 r^4+2 r^2-2 r r_+}h_1^{v\chi}
   \nonumber\\
   &&
   +\Biggl(r_+ \Bigl(-15 \mu_0 ^2 r^7
   -3 r r_+^4 (\mu_0 ^2 r^2+3)+2 r^4 r_+ (11 \mu_0 ^2 r^2+5)
   -r^3 r_+^2 (4 \mu_0 ^2 r^2+53)+6 r^5
\nonumber\\
 &&
+34 r^2 r_+^3+6 r_+^5\Bigr)
-3 \mu_2^2 r^6 (2 r-3 r_+) (r-r_+) \Biggr)
\frac{H_-'}{6 r^2
   (r-r_+) (\mu_0 ^2 r^3+2 r-2 r_+)^2} 
   \nonumber\\
   &&
   +\Biggl(r_+ \Bigl(3 r r_+^4 (\mu_0 ^2
   r^2+4)+2 r^2 r_+^3 (41-3 \mu_0 ^2 r^2)
\nonumber\\
&&   -6 r^5 (5 \mu_0 ^2 r^2+4)
   +r^4 r_+ (61 \mu_0 ^2
   r^2+64)-2 r^3 r_+^2 (11 \mu_0 ^2 r^2+76)-6 r_+^5\Bigr)
\nonumber\\
&&   -6 \mu_2^2 r^6 (r-3 r_+)
   (r-r_+)\Biggr)\frac{ H_-}{6 r^3 (r-r_+) (\mu_0 ^2 r^3+2 r-2 r_+)^2},
   \end{eqnarray}
   \begin{eqnarray}
   h^{\theta\theta}_2&=&
 \frac{1}{2 r^2}H_{2-}+\frac{2 r_+^2}{3 r^3}h_1^{v\chi}
   -\frac{r_+ (2 r-3 r_+)(3 r^2-2 r r_++r_+^2) }
             {6 r^3 (r-r_+)(\mu_0 ^2 r^3+2 r-2 r_+)}H_-'
   \nonumber\\
   &&
   +r_+ \Bigl(-3 \mu_0 ^2 r^6-2 \mu_0 ^2 r^5 r_++2 r^4 \left(\mu_0 ^2
   r_+^2-6\right)+r^3 r_+ (\mu_0 ^2 r_+^2+24)
\nonumber\\
&&-6 r^2 r_+^2+4 r r_+^3-2
   r_+^4\Bigr)
\frac{H_-}{6 r^5 (r-r_+) \left(\mu_0 ^2 r^3+2 r-2 r_+\right)}\,.
   \end{eqnarray}

\section{expressions for $\Omega\neq 0$}
\label{sec:expomn0}

\subsection{Metric perturbation at order $\epsilon^0$}

\begin{eqnarray}
h_0^{vv}&=&\frac{r  \left(2 \mu_0 ^2 r^3 (r-r_+)
+4 \Omega ^2 r^4+r_+ (r_+-2 r)\right)}{2 (r-r_+)^2
   \left(\mu_0 ^2 r^3+r_+\right)}H_- 
   \nonumber\\
   &-&\frac{r  \left(\mu_0 ^2 r (2 r (\Omega  r-1)+3 r_+)+2 \Omega  \left(2 \Omega 
   r^2+r_+\right)\right)}{\Omega  (r-r_+) \left(\mu_0 ^2 r^3+r_+\right)}
   H, 
   \\
h_0^{vr}&=&
\frac{4 \Omega ^2 r^4+r_+ (3 r_+-4 r)}{4 (r-r_+)(\mu_0 ^2r^3+r_+)}H_-
   -\frac{\mu_0 ^2 r (2 r(\Omega  r-1)+3 r_+)+2 \Omega(2 \Omega r^2+r_+)}
{2 \Omega (\mu_0 ^2 r^3+r_+)}H, 
   \\
h_0^{rr}&=&\frac{ (r-r_+) 
\left(\mu_0 ^2 (2 r-3 r_+)-4 \Omega ^2 r\right)}
{2 \Omega  \left(\mu_0 ^2 r^3+r_+\right)} H
   +\frac{ \left(4 \Omega ^2 r^4-4 r r_++3 r_+^2\right)}
{4 \mu_0 ^2 r^4+4 r r_+}H_-, 
  \\
h_0^{\theta\theta}&=&\frac{1}{2r^2}H_-\,. 
\end{eqnarray}

\subsection{Metric perturbation at order $\epsilon^1$}

\begin{eqnarray}
h^{r\chi}_1&=&
\frac{r_+}{2 \Omega r^2 (r-r_+)(\mu_0 ^2 r+4 \Omega)(\mu_0 ^2 r^3+r_+)}
   \Biggl(\mu_0 ^4 r^2 (2 r-3 r_+) (r-r_+)
   \nonumber\\
   &&
   +2 \mu_0 ^2 \Omega \Bigl(2 \Omega r^4+r_+ \left(-4
   r^2+8 r r_+-3 r_+^2\right)\Bigr)
 +4 \Omega^2 r_+ (2 \Omega r+1) (2 r-r_+)\Biggr) H
    \nonumber\\
   &&
   +\frac{r_+}{4 r^3 (r-r_+)^2 \left(\mu_0 ^2 r+4 \Omega\right) \left(\mu_0 ^2
   r^3+r_+\right)} 
 \Biggl(\mu_0^2 r^2 (r-r_+)(4 \Omega r^3-6 \Omega r r_+^2-4 r r_++3 r_+^2)
    \nonumber\\
   &&
   +2 \Omega r_+
   \Bigl(-2 r^3 \bigl(\Omega r (4 \Omega r+1)-1\bigr)
  +2 r^2 r_+ \bigl(\Omega r (2 \Omega r+1)+3\bigr)
   -13 r r_+^2+6r_+^3\Bigr)\Biggr)H_-
    \nonumber\\
   &&
   +\frac{\Omega (r_+-r) }{\mu_0 ^2 r+4 \Omega}{h^{v\chi}_1}'-\frac{\Omega 
   \left(2 \Omega r^2+r_+\right)}{r \left(\mu_0 ^2 r+4 \Omega\right)}h^{v\chi}_1. 
\end{eqnarray}

\subsection{Metric perturbation at order $\epsilon^2$}

 \begin{eqnarray}
   h_2^{rr}&=&
   \frac{2(r-r_+)
   \Bigl(2r\bigl(r\mu_0 ^2+\Omega(\mu_0 ^2r^2+2\Omega r+4)\bigr)
   -(r \mu_0 ^2+6 \Omega) r_+\Bigr) r_+^2}
   {r^2 (r \mu_0 ^2+4 \Omega ) (\mu_0 ^2 r^3+r_+)} h_1^{v\chi}
   \nonumber\\
   &&
   +\frac{2(r \mu_0 ^2+2 \Omega)(r-r_+)^2 r_+^2}
   {r (r \mu_0 ^2+4 \Omega )(\mu_0 ^2 r^3+r_+)} {h_1^{v\chi}}'
   \nonumber\\
   &&
  +\frac{ 4 \Omega ^2 r^4-4 r_+ r+3 r_+^2}{4 \mu_0 ^2 r^4+4 r_+ r}H_{2-}
   -\frac{ (r-r_+) \Bigl((3 r_+-2 r) \mu_0 ^2+4 \Omega ^2 r\Bigr)}
  {2 \Omega  \left(\mu_0 ^2r^3+r_+\right)}H_2
   \nonumber\\
   &&
   -\frac{1}{24\Omega r^4(r\mu_0 ^2+4\Omega)(r-r_+)(\mu_0 ^2 r^3+r_+)^2}
    \Biggl(6 \mu_2^2 \Omega  (r \mu_0 ^2+4 \Omega ) (r-r_+)
    \bigl(4\Omega ^2 r^4+r_+ (3 r_+-4 r)\bigr) r^6
   \nonumber\\
   &&
   +r_+ \Biggl[r^3 (r-r_+) \Bigl(24 \Omega  \bigl(\Omega  r (\Omega 
   r-1)+1\bigr) r^5
  +4 \bigl(\Omega  r (\Omega  r (2 \Omega  r+3)-15)+6\bigr) r_+ r^3
   \nonumber\\
   &&
   -2 \bigl(\Omega  r (6 \Omega  r-17)+15\bigr) r_+^2 r^2-3
   (2 \Omega  r+9) r_+^3 r+3 (9-16 \Omega  r) r_+^4\Bigr) \mu_0 ^4
   \nonumber\\
   &&
   +\Omega  r \Bigl(24 \Omega  (\Omega  r (2 \Omega  r
   (\Omega  r+3)-3)+4) r^7
 +16 (\Omega  r (\Omega  r (2 \Omega  r (\Omega  r-2)+3)-24)+6) r_+ r^5
   \nonumber\\
   &&
   +2 (2 \Omega  r (4 \Omega 
   r (\Omega  r (\Omega  r-5)-8)+79)-105) r_+^2 r^4
  -4 (\Omega  r (19 \Omega  r+16)-42) r_+^3 r^3
   \nonumber\\
   &&
   +3 (4 \Omega  r (7
   \Omega  r-1)+39) r_+^4 r^2
 +3 (16 \Omega  r-81) r_+^5 r+102 r_+^6\Bigr) \mu_0 ^2
 +4 \Omega ^2 (24 \Omega^2 (2 \Omega  r (\Omega  r+1)+1) r^8
   \nonumber\\
   &&
   +16 \Omega  (2 \Omega ^3 r^3-6 \Omega  r-3) r_+ r^6+2 (2 \Omega  r (4
   \Omega  r (\Omega  r (\Omega  r-3)-2)+9)+3) r_+^2 r^4
   \nonumber\\
   &&
   +4 (\Omega  r (6-7 \Omega  r)+9) r_+^3 r^3+3 (2 \Omega  r-1)
   (6 \Omega  r+1) r_+^4 r^2-51 r_+^5 r+30 r_+^6)\Biggr]\Biggr)H_-
   \nonumber\\
   &&
   +\frac{1}{12 \Omega ^2 r^3 (r \mu_0 ^2+4 \Omega ) (\mu_0 ^2 r^3+r_+)^2} 
   \Biggr(6 \mu_2^2 \Omega  (r \mu_0 ^2+4 \Omega) (r-r_+)
      \bigl(4 \Omega ^2 r^4+(2 r-3 r_+) r_+\bigr) r^3
   \nonumber\\
   &&
   +r_+ \Biggl[-3 r^3 (r-r_+)
\Bigl(10 \Omega  r^4-2 (\Omega  r_++2) r^3+2 r_+ (4-3 \Omega  r_+) r^2
   +r_+^2 (3-2 \Omega  r_+) r-9 r_+^3\Bigr) \mu_0 ^6
   \nonumber\\
   &&
    +\Omega  r \Bigl(12 (2 \Omega ^3 r^3-14 \Omega  r+3) r^5
  +4 (\Omega  r (\Omega  r (6
   \Omega  r+5)+48)-30) r_+ r^4
   \nonumber\\
   &&
   +2 (\Omega  r (2 \Omega  r+21)+50) r_+^2 r^3+(83-102 \Omega  r) r_+^3 r^2-3 (8
   \Omega  r+59) r_+^4 r+48 r_+^5\Bigr) \mu_0 ^4
   \nonumber\\
   &&
   +2 \Omega ^2 \Bigl(12 (\Omega  r (2 \Omega  r (3 \Omega  r+1)-7)-2)
   r^5+4 (\Omega  r (\Omega  r (16 \Omega  r+13)+18)+12) r_+ r^4
   \nonumber\\
   &&
   +2 \bigl(\Omega  r (2 \Omega  r (2 \Omega  r-1)+13)-14\bigr)
   r_+^2 r^3-(176 \Omega  r+5) r_+^3 r^2
   +3 (10 \Omega  r-27) r_+^4 r+24 r_+^5\Bigr) \mu_0 ^2
   \nonumber\\
   &&
   +8 \Omega ^3 \Bigl(12 \Omega  (2 \Omega  r (\Omega  r+1)+1) r^5
  +4 \Omega  (\Omega  r (4 \Omega  r+3)-6) r_+ r^4
   \nonumber\\
   &&
   +2 \bigl(\Omega  r (2\Omega  r (2 \Omega  r-3)-5)-6\bigr) r_+^2 r^2
-(20 \Omega  r+3) r_+^3 r+3 (2 \Omega  r-1)
   r_+^4\Bigr)\Biggr]\Biggr)H,  
   \end{eqnarray}

   \begin{eqnarray}
   h_2^{vv}&=&
   \frac{4  r (r \mu_0 ^2+2 \Omega ) r_+^2}{(r \mu_0 ^2+4 \Omega ) (\mu_0 ^2
   r^3+r_+)}{h_1^{v\chi}}'
   +\frac{ 8 r (r \mu_0 ^2+\Omega  (\mu_0 ^2 r^2+2 \Omega  r+4))r_+^2
   -4 (r \mu_0 ^2+6 \Omega ) r_+^3}
   {(r \mu_0 ^2+4 \Omega ) (r-r_+) (\mu_0 ^2
   r^3+r_+)}
   h_1^{v\chi}
   \nonumber\\
   &&
   +\frac{ r (4 \Omega ^2 r^4+2 \mu_0 ^2 (r-r_+) r^3+r_+ (r_+-2 r))}{2
   (r-r_+)^2 (\mu_0 ^2 r^3+r_+)}
   H_{2-}
\nonumber\\
&&
   +\frac{ r}{r-r_+}\left(\frac{r (\mu_0 ^2 (2 r-3 r_+)-4 \Omega ^2
   r)}{\Omega  (\mu_0 ^2 r^3+r_+)}-2\right) H_2
   \nonumber\\
   &&
   -\frac{ 1}{12 \Omega  r^2 (r \mu_0 ^2+4 \Omega ) (r-r_+)^3 (\mu_0 ^2
   r^3+r_+)^2}
\nonumber\\
&&\times  
 \Biggl(6 \mu_2^2 \Omega (r \mu_0 ^2+4 \Omega ) (r-r_+)
   (4 \Omega ^2 r^4+r_+ (3 r_+-4 r)) r^6
   \nonumber\\
   &&
   +r_+ \Biggl[r^3(r-r_+) \Bigl(24 \Omega  (\Omega  r (\Omega  r-1)+1) r^5
 +4 (\Omega  r (\Omega  r (2 \Omega  r+3)-15)+6) r_+ r^3
   \nonumber\\
   &&
   -2
   (\Omega  r (6 \Omega  r-17)+15) r_+^2 r^2
  -3 (2 \Omega  r+9) r_+^3 r+3 (9-16 \Omega  r) r_+^4\Bigr) \mu_0^4
   \nonumber\\
   &&
   +\Omega  r (24 \Omega  (\Omega  r (2 \Omega  r (\Omega  r+3)-3)+4) r^7+16 (\Omega  r (\Omega  r (2 \Omega  r (\Omega 
   r-2)+3)-24)+6) r_+ r^5
   \nonumber\\
   &&
   +2 (2 \Omega  r (4 \Omega  r (\Omega  r (\Omega  r-5)-8)+79)-105) r_+^2 r^4
   -4 (\Omega  r (19\Omega  r+16)-42) r_+^3 r^3
   \nonumber\\
   &&
   +3 (4 \Omega  r (7 \Omega  r-1)+39) r_+^4 r^2+3 (16 \Omega  r-81) r_+^5 r+102
   r_+^6) \mu_0 ^2+4 \Omega ^2 (24 \Omega ^2 (2 \Omega  r (\Omega  r+1)+1) r^8
   \nonumber\\
   &&
   +16 \Omega  (2 \Omega ^3 r^3-6
   \Omega  r-3) r_+ r^6
  +2 (2 \Omega  r (4 \Omega  r (\Omega  r (\Omega  r-3)-2)+9)+3) r_+^2 r^4
   \nonumber\\
   &&
   +4 (\Omega  r
   (6-7 \Omega  r)+9) r_+^3 r^3+3 (2 \Omega  r-1) (6 \Omega  r+1) r_+^4 r^2
  -51 r_+^5 r+30 r_+^6)\Biggr]\Biggr)H_-
   \nonumber\\
   &&
 +\frac{1}{6 \Omega ^2 r(r \mu_0^2+4 \Omega )(r-r_+)^2 (\mu_0 ^2 r^3+r_+)^2}
\nonumber\\
&&\times 
   \Biggl(6 \mu_2^2 \Omega  (r \mu_0 ^2+4 \Omega ) (r-r_+) (4 \Omega ^2
   r^4+(2 r-3 r_+) r_+) r^3
   \nonumber\\
   &&
   +r_+ (-3 r^3 (r-r_+) (10 \Omega  r^4-2 (\Omega  r_++2)
   r^3+2 r_+ (4-3 \Omega  r_+) r^2+r_+^2 (3-2 \Omega  r_+) r-9 r_+^3) \mu_0 ^6
   \nonumber\\
   &&
   +\Omega  r
   (12 (2 \Omega ^3 r^3-14 \Omega  r+3) r^5+4 (\Omega  r (\Omega  r (6 \Omega  r+5)+48)-30) r_+ r^4
   \nonumber\\
   &&
   +2
   (\Omega  r (2 \Omega  r+21)+50) r_+^2 r^3+(83-102 \Omega  r) r_+^3 r^2-3 (8 \Omega  r+59) r_+^4 r+48
   r_+^5) \mu_0 ^4
   \nonumber\\
   &&
   +2 \Omega ^2 (12 (\Omega  r (2 \Omega  r (3 \Omega  r+1)-7)-2) r^5+4 (\Omega  r (\Omega  r (16
   \Omega  r+13)+18)+12) r_+ r^4
   \nonumber\\
   &&
   +2 (\Omega  r (2 \Omega  r (2 \Omega  r-1)+13)-14) r_+^2 r^3-(176 \Omega  r+5)
   r_+^3 r^2+3 (10 \Omega  r-27) r_+^4 r+24 r_+^5) \mu_0 ^2
   \nonumber\\
   &&
   +8 \Omega ^3 (12 \Omega  (2 \Omega  r
   (\Omega  r+1)+1) r^5+4 \Omega  (\Omega  r (4 \Omega  r+3)-6) r_+ r^4
   \nonumber\\
   &&
   +2 (\Omega  r (2 \Omega  r (2 \Omega  r-3)-5)-6)
   r_+^2 r^2-(20 \Omega  r+3) r_+^3 r+3 (2 \Omega  r-1) r_+^4))\Biggr)H,  
   \end{eqnarray}
   \begin{eqnarray}
   h_2^{vr}
   &=&
   \frac{2  (r \mu_0 ^2+2 \Omega ) (r-r_+) r_+^2}
   {(r \mu_0 ^2+4 \Omega ) (\mu_0 ^2r^3+r_+)}
   {h_1^{v\chi}}'
   +\frac{ 4 r (r \mu_0 ^2+\Omega (\mu_0 ^2 r^2+2 \Omega  r+4))r_+^2
    -2 (r \mu_0 ^2+6 \Omega ) r_+^3}{r (r \mu_0 ^2+4 \Omega )
   (\mu_0 ^2 r^3+r_+)}h_1^{v\chi}
   \nonumber\\
   &&
   +\frac{1}{2} \left(\frac{r (\mu_0 ^2 (2 r-3 r_+)-4 \Omega ^2 r)}{\Omega 
   (\mu_0 ^2 r^3+r_+)}-2\right) H_2
   +\frac{4\Omega^2r^4+r_+(3 r_+-4 r)}{4(r-r_+)(\mu_0^2r^3+r_+)}H_{2-}
   \nonumber\\
   &&
   -\frac{1}{24 \Omega  r^3(r \mu_0 ^2+4 \Omega )
(r-r_+)^2 (\mu_0 ^2 r^3+r_+)^2}
\nonumber\\
&&\times\Biggl(6 \mu_2^2 \Omega  (r \mu_0 ^2+4 \Omega )
   (r-r_+) (4 \Omega ^2 r^4
   +r_+ (3 r_+-4 r)) r^6
   \nonumber\\
   &&
   +r_+ \Biggl[r^3 (r-r_+) \Bigl(24 \Omega(\Omega  r (\Omega  r-1)+1) r^5
   +4 (\Omega  r (\Omega  r (2 \Omega  r+3)-15)+6) r_+ r^3
   \nonumber\\
   &&
   -2 (\Omega  r (6 \Omega  r-17)+15)
   r_+^2 r^2-3 (2 \Omega  r+9) r_+^3 r+3 (9-16 \Omega  r) r_+^4\Bigr) \mu_0 ^4
   \nonumber\\
   &&
   +\Omega  r \Bigl(24 \Omega 
   (\Omega  r (2 \Omega  r (\Omega  r+3)-3)+4) r^7
  +16 (\Omega  r (\Omega  r (2 \Omega  r (\Omega  r-2)+3)-24)+6) r_+ r^5
   \nonumber\\
   &&
   +2
   (2 \Omega  r (4 \Omega  r (\Omega  r (\Omega  r-5)-8)+79)-105) r_+^2 r^4
  -4 (\Omega  r (19 \Omega  r+16)-42) r_+^3r^3
   \nonumber\\
   &&
   +3 (4 \Omega  r (7 \Omega  r-1)+39) r_+^4 r^2
   +3 (16 \Omega  r-81) r_+^5 r+102 r_+^6\Bigr) \mu_0 ^2
\nonumber\\
&&
  +4\Omega ^2 \Bigl(24 \Omega ^2 (2 \Omega  r (\Omega  r+1)+1) r^8
  +16 \Omega  (2 \Omega ^3 r^3-6 \Omega  r-3) r_+ r^6
   \nonumber\\
   &&
  +2\bigl(2\Omega  r(4\Omega  r(\Omega  r(\Omega r-3)-2)+9)+3\bigr)r_+^2r^4
   +4 (\Omega  r (6-7 \Omega  r)+9) r_+^3r^3
   \nonumber\\
   &&
  +3 (2 \Omega  r-1) (6 \Omega  r+1) r_+^4 r^2-51 r_+^5 r+30 r_+^6\Bigr)
\Biggr]
 \Biggr) H_-
   \nonumber\\
   &&
   +\frac{1}{12\Omega ^2r^2(r\mu_0 ^2+4\Omega)(r-r_+)(\mu_0^2r^3+r_+)^2}
\nonumber\\
&&\times \Biggl(6 \mu_2^2 \Omega 
   (r \mu_0 ^2+4 \Omega ) (r-r_+) (4 \Omega ^2 r^4+(2 r-3 r_+) r_+) r^3
   \nonumber\\
   &&
   +r_+ \Biggl[-3r^3 (r-r_+)\Bigl(10 \Omega  r^4-2 (\Omega  r_++2) r^3
\nonumber\\
&& +2 r_+ (4-3 \Omega  r_+) r^2+r_+^2 (3-2\Omega  r_+) r-9 r_+^3\Bigr) \mu_0 ^6
   \nonumber\\
   &&
   +\Omega  r \Bigl(12 (2 \Omega ^3 r^3-14 \Omega  r+3) r^5+4 (\Omega 
   r (\Omega  r (6 \Omega  r+5)+48)-30) r_+ r^4
   \nonumber\\
   &&
   +2 (\Omega  r (2 \Omega  r+21)+50) r_+^2 r^3+(83-102 \Omega  r)
   r_+^3 r^2-3 (8 \Omega  r+59) r_+^4 r+48 r_+^5\Bigr) \mu_0 ^4
   \nonumber\\
   &&
   +2 \Omega ^2 \Bigl(12 (\Omega  r(2\Omega r(3\Omega  r+1)-7)-2) r^5
   +4 (\Omega  r (\Omega  r (16 \Omega  r+13)+18)+12) r_+ r^4
   \nonumber\\
   &&
   +2 (\Omega  r (2 \Omega  r (2 \Omega 
   r-1)+13)-14) r_+^2 r^3
\nonumber\\
&&-(176 \Omega  r+5) r_+^3 r^2
   +3 (10 \Omega  r-27) r_+^4 r+24 r_+^5\Bigr) \mu_0^2
   \nonumber\\
   &&
   +8 \Omega ^3 \Bigl(12 \Omega  (2 \Omega  r (\Omega  r+1)+1) r^5
   +4 \Omega  (\Omega  r (4 \Omega  r+3)-6) r_+ r^4
   \nonumber\\
   &&
   +2(\Omega  r (2 \Omega  r (2 \Omega  r-3)-5)-6) r_+^2 r^2
   -(20 \Omega  r+3) r_+^3 r+3 (2 \Omega  r-1) r_+^4\Bigr)
   \Biggr]\Biggr)H, 
   \end{eqnarray}
   \begin{eqnarray}
   h_2^{\theta\theta}&=&
   \frac{2  \Omega  r_+ (r_+-r)}{3 r^2 (\mu_0 ^2 r+4 \Omega )}{h_1^{v\chi}}'
   +\frac{2r_+(\mu_0 ^2 r r_+-2 \Omega ^2 r^2+3 \Omega  r_+)}
    {3 r^3 (\mu_0 ^2 r+4 \Omega )} h_1^{v\chi} 
   +\frac{1}{2 r^2}H_{2-}
   \nonumber\\
   &&
   +\frac{r_+}{6\Omega r^4(r-r_+) (\mu_0 ^2 r+4 \Omega)(\mu_0^2r^3+r_+)}
\nonumber\\
&&\times
 \Biggl(\mu_0^4 r^2
   \Bigl(6 r^3 (\Omega  r-1)-2 r r_+^2 (\Omega  r+4)+13 r^2 r_++3 r_+^3\Bigr)
   \nonumber\\
   &&
   +2 \mu_0^2 \Omega \Bigl(6 r^4 (3\Omega  r-2)+r^3 r_+ (4 \Omega r+21)
  -2 r^2 r_+^2 (3 \Omega  r+2)+9 r r_+^3-6 r_+^4\Bigr)
   \nonumber\\
   &&
   +8 \Omega^2\Bigl(6 \Omega  r^4-2 r_+^3 (\Omega  r+1)
 +2 r r_+^2 (\Omega  r+1)+3 r^2 r_+\Bigr)
 \Biggr) H
   \nonumber\\
   &&
   +\frac{ r_+}{12 r^5 (r-r_+)^2(\mu_0 ^2 r+4 \Omega)(\mu_0^2 r^3+r_+)}
\nonumber\\
 &&\times
   \Biggl(-2 \mu_0 ^4 r^4(r-r_+) (r+r_+) (3 r^2-r r_+-r_+^2)
   \nonumber\\
   &&
   +\mu_0^2 r \Bigl(-12 \Omega  r^6 (\Omega  r+2)+2 r^4
   r_+ (8 \Omega  r+3)
  +r^3 r_+^2 (4 \Omega  r (\Omega  r+6)-15)+2 r^2 r_+^3 (9-10 \Omega  r)
   \nonumber\\
   &&
   +r r_+^4 (4\Omega  r-5)-2 r_+^5\Bigr)
  +4 \Omega \Bigl(-12 \Omega ^2 r^7
  -r^3 r_+^2 (2 \Omega  r (2 \Omega  r+1)+5)
   \nonumber\\
   &&
   +2 r^2
   r_+^3 (\Omega  r (2 \Omega  r+1)+5)+6 r^4 r_+-12 r r_+^4+4 r_+^5\Bigr)
 \Biggr)H_-. 
   \end{eqnarray}
   
\subsection{$A_H$, $A_{H_-}$, $B_H$, $B_{H_-}$, $C_H$ and $C_{H_-}$}

\begin{eqnarray}
&&A_H(r;r_+,\mu_0,\Omega)
\nonumber\\
&&:=
\frac{r_+}{2 \Omega r^3 (r-r_+)^3 \left(\mu_0 ^2
   r+4 \Omega\right) \left(\mu_0 ^2 r^3+r_+\right)}  
   \Bigl(\mu_0 ^6 r^4 r_+ (r-3 r_+) (r-r_+)
   \nonumber\\
   &&
   +2 \mu_0 ^4 r \bigl(-2 r^4 (\Omega r (\Omega
   r+1)-3)-r^3 r_+ (\Omega r (4 \Omega r+1)+6)
   \nonumber\\
   &&
   +2 r^2 r_+^2 (\Omega r-7) (\Omega r+1)+9 r r_+^3
   (\Omega r+2)-6 r_+^4\bigr)
   \nonumber\\
   &&
   -8 \mu_0 ^2 \Omega \bigl(r^4 (\Omega r+1) (3 \Omega r-2)
   +r^3 r_+ (\Omega r
   (6 \Omega r+11)-3)
   \nonumber\\
   &&
   +r^2 r_+^2 (19-\Omega r (3 \Omega r+2))-r r_+^3 (\Omega r+14)+3
   r_+^4\bigr)
   \nonumber\\
   &&
   +16 \Omega^2 \left(-2 \Omega r^4 (\Omega r+1)-4 \Omega r^3 r_+ (\Omega r+2)+r_+^3
   (3-2 \Omega r)+r r_+^2 (2 \Omega r (\Omega r+3)-5)\right)\Bigr),~~~~~~~~
\end{eqnarray}
\begin{eqnarray}
&&A_{H_-}(r;r_+,\mu_0,\Omega):=
\nonumber\\
&&\frac{r_+}{4 r^4 (r-r_+)^4 \left(\mu_0 ^2 r+4 \Omega\right) \left(\mu_0 ^2
   r^3+r_+\right)}  \Bigl(\mu_0 ^4 r^4 (r-r_+) \bigl(8
   \Omega r^4-2 r^3 (\Omega r_++8)
   \nonumber\\
   &&
   +2 r^2 r_+ (3 \Omega r_++13)+17 r r_+^2-21
   r_+^3\bigr)
   \nonumber\\
   &&
   +4 \mu_0 ^2 r \bigl(2 \Omega r^6 (\Omega r-1) (\Omega r+6)
   +r^4 r_+ (\Omega r (\Omega r (4
   \Omega r+1)+34)-10)
   \nonumber\\
   &&
   -2 r^3 r_+^2 (\Omega r (\Omega r (\Omega r+3)+8)-6)-r^2 r_+^3 (\Omega r
   (\Omega r+19)-10)+2 r r_+^4 (6 \Omega r-11)+9 r_+^5\bigr)
   \nonumber\\
   &&
   +8 \Omega \bigl(4 \Omega^2 r^7 (\Omega
   r+1)+2 r^4 r_+ (\Omega r (\Omega r (4 \Omega r+7)-1)-5)
   \nonumber\\
   &&
   -2 r^3 r_+^2 (\Omega r+1) (2 \Omega r
   (\Omega r+3)-3)+3 r^2 r_+^3 (2 \Omega r (\Omega r+2)+7)
   \nonumber\\
   &&
   -r r_+^4 (6 \Omega r+31)+12
   r_+^5\bigr)\Bigr), 
   \end{eqnarray}
   \begin{eqnarray}
   &&B_H(r;r_+,\mu_0,\mu_2,\Omega):=
   \nonumber\\
   &&\frac{1}{6 \Omega ^2 r^3
   (r \mu_0 ^2+4 \Omega ) (r-r_+) (\mu_0 ^2 r^3+r_+)^2} (6 \mu_2^2 \Omega 
   (r \mu_0 ^2+4 \Omega ) (r-r_+) (\mu_0 ^4 r^6+4 \Omega ^2 r^4
   \nonumber\\
   &&
   +2 \mu_0 ^2 r_+ r^3+2 (r-r_+)
   r_+) r^3+r_+ (r^6 (r-r_+) (6 (\Omega  r+1) r^2+(2 \Omega  r-3) r_+ r-9
   r_+^2) \mu_0 ^8
   \nonumber\\
   &&
   +r^3 (6 (\Omega  r (2 \Omega  r (\Omega  r+2)-1)+2) r^4+2 (\Omega  r+3) (2 \Omega  r (2 \Omega 
   r-1)-5) r_+ r^3
   \nonumber\\
   &&
   +(\Omega  r (4 \Omega  r (\Omega  r+1)+21)+6) r_+^2 r^2+(43 \Omega  r+30) r_+^3 r-18 (\Omega 
   r+1) r_+^4) \mu_0 ^6
   \nonumber\\
   &&
   +4 \Omega  r (3 (8 \Omega ^3 r^3-14 \Omega  r+3) r^5+4 (\Omega  r (\Omega  r (5
   \Omega  r+16)+8)-6) r_+ r^4
   \nonumber\\
   &&
   +(2 \Omega  r (\Omega  r (2 \Omega  r+11)+30)+9) r_+^2 r^3-2 (\Omega  r-2) (\Omega 
   r+6) r_+^3 r^2-6 (3 \Omega  r+5) r_+^4 r+9 r_+^5) \mu_0 ^4
   \nonumber\\
   &&
   +4 \Omega ^2 r (6 (\Omega  r (2 \Omega  r
   (5 \Omega  r+1)-7)-2) r^4+4 (\Omega  r (2 \Omega  r+3) (7 \Omega  r+5)+6) r_+ r^3
   \nonumber\\
   &&
   +(\Omega  r (2 \Omega  r+9) (2 \Omega 
   r+11)-36) r_+^2 r^2+(42-\Omega  r (8 \Omega  r+87)) r_+^3 r+12 (\Omega  r-2) r_+^4) \mu_0 ^2
   \nonumber\\
   &&
   +16 \Omega
   ^3 (6 \Omega  (2 \Omega  r (\Omega  r+1)+1) r^5
   \nonumber\\
   &&
   +4 \Omega  (\Omega  r (2 \Omega  r+3)-3) r_+ r^4+\Omega  (4
   \Omega ^2 r^2+15) r_+^2 r^3-3 (\Omega  r-2) r_+^3 r-6 r_+^4))), 
   \end{eqnarray}
   \begin{eqnarray}
   &&B_{H_-}(r;r_+,\mu_0,\mu_2,\Omega):=
   \nonumber\\
   &&\frac{1}{12 \Omega  r^4 (r
   \mu_0 ^2+4 \Omega ) (r-r_+)^2 (\mu_0 ^2 r^3+r_+)^2}
    (r_+ (-3
   \mu_0 ^6 (r-r_+) (8 \Omega ^2 (r+r_+) r^4
   \nonumber\\
   &&
   +2 \Omega  (-3 r^3+3 r_+ r^2+3 r_+^2
   r+r_+^3) r+(4 r-3 r_+) r_+ (r+r_+)) r^6
   \nonumber\\
   &&
   +\mu_0 ^4 (-24 \Omega  (\Omega  r (\Omega  r
   (\Omega  r+5)-4)+1) r^6-4 (\Omega  r (\Omega  r (\Omega  r (6 \Omega  r+13)+54)-18)+6) r_+ r^4
   \nonumber\\
   &&
   +2 (\Omega  r (3 \Omega  r
   (14 \Omega  r+5)-31)+21) r_+^2 r^3+(\Omega  r (2 \Omega  r (8 \Omega  r+39)+57)+6) r_+^3 r^2
   \nonumber\\
   &&
   +(\Omega  r (24 \Omega
    r-1)-42) r_+^4 r+6 (3-4 \Omega  r) r_+^5) r^3+2 \mu_0 ^2 \Omega  (-12 \Omega  (3 \Omega  r (2 \Omega  r
   (\Omega  r+1)-1)+4) r^7
   \nonumber\\
   &&
   -4 (\Omega  r (\Omega  r (\Omega  r (16 \Omega  r+29)+24)-66)+12) r_+ r^5
   \nonumber\\
   &&
   -2 (\Omega  r (\Omega  r
   (2 \Omega  r (2 \Omega  r+3)-21)+86)-45) r_+^2 r^4+(\Omega  r (16 \Omega  r (2 \Omega  r+7)-27)-50) r_+^3
   r^3
   \nonumber\\
   &&
   +(\Omega  r (1-34 \Omega  r)-50) r_+^4 r^2+2 (12 \Omega  r+47) r_+^5 r-42 r_+^6) r+8 \Omega ^2
   (-12 \Omega ^2 (2 \Omega  r (\Omega  r+1)+1) r^8
   \nonumber\\
   &&
   +4 \Omega  (\Omega  r (12-\Omega  r (4 \Omega  r+3))+6) r_+ r^6-2
   (\Omega  r (\Omega  r (2 \Omega  r (2 \Omega  r-3)+9)+12)-3) r_+^2 r^4
   \nonumber\\
   &&
   +(\Omega  r (4 \Omega  r-9)-2) r_+^3 r^3+(5
   \Omega  r (3-2 \Omega  r)-14) r_+^4 r^2+16 r_+^5 r-6 r_+^6))
   \nonumber\\
   &&
   -6 \mu_2^2 \Omega  r^6 (r
   \mu_0 ^2+4 \Omega ) (r-r_+) (4 \Omega ^2 r^4+r_+ (3 r_+-4 r))), 
   \end{eqnarray}
   \begin{eqnarray}
   &&C_H(r;r_+,\mu_0,\mu_2,\Omega):=
   \nonumber\\
   &&\frac{1}{12 \Omega  r^3 (r \mu_0 ^2+4 \Omega )
   (r-r_+)^2 (\mu_0 ^2 r^3+r_+)^2} 
   (6 \mu_2^2 \Omega  (r \mu_0 ^2+4 \Omega ) (r-r_+) (4 \Omega ^2
   r^4+(2 r-3 r_+) r_+) r^5
   \nonumber\\
   &&
   +r_+ (-r^5 (r-r_+) (6 (7 \Omega  r-4) r^3-2 (\Omega  r-15)
   r_+ r^2+(51-10 \Omega  r) r_+^2 r-3 (2 \Omega  r+21) r_+^3) \mu_0 ^6
   \nonumber\\
   &&
   +r^2 (12 \Omega  (2 \Omega
   ^3 r^3-20 \Omega  r+7) r^6+4 (\Omega  r (\Omega  r (\Omega  r (6 \Omega  r+5)+56)-54)+3) r_+ r^4
   \nonumber\\
   &&
   +2 (\Omega  r
   (\Omega  r (2 \Omega  r+1)+34)-9) r_+^2 r^3+(\Omega  r (243-70 \Omega  r)-36) r_+^3 r^2\nonumber\\
   &&
   +(78-\Omega  r (24 \Omega 
   r+353)) r_+^4 r+12 (10 \Omega  r-3) r_+^5) \mu_0 ^4+2 \Omega  (12 \Omega  (\Omega  r (2 \Omega  r (3
   \Omega  r+1)-11)-2) r^7
   \nonumber\\
   &&
   +4 (\Omega  r (\Omega  r (\Omega  r (16 \Omega  r+13)+18)-3)+6) r_+ r^5
   \nonumber\\
   &&
   +2 (\Omega  r (\Omega  r
   (2 \Omega  r (2 \Omega  r-1)+37)+2)-21) r_+^2 r^4-(\Omega  r (272 \Omega  r+81)+20) r_+^3 r^3
   \nonumber\\
   &&
   +(\Omega  r (78
   \Omega  r-25)+82) r_+^4 r^2+4 (6 \Omega  r-23) r_+^5 r+36 r_+^6) \mu_0 ^2+8 \Omega ^2 (12 \Omega ^2
   (2 \Omega  r (\Omega  r+1)+1) r^7
   \nonumber\\
   &&
   +4 \Omega  (\Omega  r (\Omega  r (4 \Omega  r+3)-6)-3) r_+ r^5+2 (\Omega  r (\Omega  r
   (2 \Omega  r (2 \Omega  r-3)-5)-6)-3) r_+^2 r^3
   \nonumber\\
   &&
   +(\Omega  r (9-20 \Omega  r)+4) r_+^3 r^2+(3 \Omega  r (2 \Omega 
   r-9)-14) r_+^4 r+2 (6 \Omega  r+5) r_+^5))), 
   \end{eqnarray}
   \begin{eqnarray}
   &&C_{H_-}(r;r_+,\mu_0,\mu_2,\Omega):=
   \nonumber\\
   &&\frac{1}{24 r^4 (r \mu_0 ^2+4 \Omega ) (r-r_+)^3 (\mu_0 ^2
   r^3+r_+)^2} (r_+ (-12 \mu_0 ^6 (2 r-3 r_+)
   (r-r_+)^2 (r+r_+) r^8
   \nonumber\\
   &&
   -\mu_0 ^4 (r-r_+) (24 \Omega  (\Omega  r (\Omega  r-2)+5) r^5+4 (\Omega  r (\Omega 
   r (2 \Omega  r-3)-51)+24) r_+ r^3
   \nonumber\\
   &&
   -2 (\Omega  r (6 \Omega  r+7)+48) r_+^2 r^2+3 (46 \Omega  r-47) r_+^3 r+15
   (9-8 \Omega  r) r_+^4) r^5
   \nonumber\\
   &&
   +\mu_0 ^2 (-24 \Omega ^2 (\Omega  r+4) (2 \Omega  r (\Omega  r-1)+1) r^8-8 \Omega 
   (\Omega  r (2 \Omega  r (2 \Omega  r (\Omega  r-2)+3)-51)+48) r_+ r^6
   \nonumber\\
   &&
   +2 (\Omega  r (2 \Omega  r (4 \Omega  r (\Omega  r
   (5-\Omega  r)+2)-79)+381)-24) r_+^2 r^4
   \nonumber\\
   &&
   +2 (2 \Omega  r (\Omega  r (67 \Omega  r+22)-66)+39) r_+^3 r^3-3 (3 \Omega 
   r (4 \Omega  r (5 \Omega  r+1)+85)-20) r_+^4 r^2
   \nonumber\\
   &&
   -3 (\Omega  r (16 \Omega  r-305)+54) r_+^5 r+6 (12-53 \Omega  r)
   r_+^6) r^2+4 \Omega  (-24 \Omega ^3 (2 \Omega  r (\Omega  r+1)+1) r^{10}
   \nonumber\\
   &&
   +8 \Omega ^2 (-4 \Omega ^3
   r^3+12 \Omega  r+9) r_+ r^8-2 (\Omega  r (2 \Omega  r (4 \Omega  r (\Omega  r (\Omega  r-3)-2)+9)+3)+24)
   r_+^2 r^5
   \nonumber\\
   &&
   +3 \Omega  (4 \Omega  r (5-3 \Omega  r)+5) r_+^4 r^4+2 (2 \Omega  r (\Omega  r (7 \Omega  r-12)-9)+39)
   r_+^3 r^4
   \nonumber\\
   &&
   -3 (\Omega  r (8 \Omega  r-13)+38) r_+^5 r^2+6 (19-5 \Omega  r) r_+^6 r-36
   r_+^7))
   \nonumber\\
   &&
   -6 \mu_2^2 \Omega  r^8 (r \mu_0 ^2+4 \Omega ) (r-r_+) (4 \Omega ^2
   r^4+r_+ (3 r_+-4 r))). 
   \end{eqnarray}
   
\subsection{Expansion of master variables near horizon for $\Omega\neq0$}
   
   \begin{eqnarray}
   H&=&\frac{\Omega  r_+ (2 \Omega  r_+-1)}{(r-r_+) \left(2 \mu_0 ^2 r_++4 \Omega \right)}+\frac{\mu_0 ^2 \Omega 
   r_+^2+\Omega }{2 \mu_0 ^2 r_++4 \Omega }
   \nonumber\\
   &&
   -\frac{\Omega  (r-r_+) \left(-2 \Omega  r_+ \left(\mu_0 ^4
   r_+^4-7\right)-3 \mu_0 ^4 r_+^4+8 \mu_0 ^2 r_+^2+4 \Omega ^2 r_+^2+4\right)}{8 r_+ (\Omega 
   r_++1) (2 \Omega  r_++1) \left(\mu_0 ^2 r_++2 \Omega \right)}
   +\mathcal O((r-r_+)^2), 
   \\
   H_-&=&1+
   \frac{(r-r_+) \left(\mu_0 ^4 r_+^3-4 \mu_0 ^2 r_+-8 \Omega ^2 r_+-6 \Omega \right)}{r_+ (2 \Omega 
   r_++1) \left(\mu_0 ^2 r_++2 \Omega \right)}
   \nonumber\\
   &&
   +\frac{(r-r_+)^2}{4 r_+^2 (\Omega  r_++1) (2 \Omega 
   r_++1) \left(\mu_0 ^2 r_++2 \Omega \right)}
    (\Omega ^2 \left(108 r_+-8 \mu_0 ^2
   r_+^3\right)
   +\Omega  \left(-2 \mu_0 ^4 r_+^4+34 \mu_0 ^2 r_+^2+48\right)
   \nonumber\\
   &&
   +\mu_0 ^2 r_+ \left(\mu_0 ^4
   r_+^4-8 \mu_0 ^2 r_+^2+32\right)+48 \Omega ^3 r_+^2)+\mathcal O((r-r_+)^2), 
   \end{eqnarray}
   \begin{eqnarray}
   h_1^{v\chi}&=&
   \beta+
   \frac{(r-r_+)}{4 r_+^2 (\Omega  r_++1)^2 (2 \Omega 
   r_++1) (2 \Omega  r_++3) (\mu_0 ^2 r_++2 \Omega ) (\mu_0 ^2 r_++4 \Omega )} 
   \nonumber\\
   &&
   (8 \beta  \mu_0 ^6 \Omega ^3 r_+^8-4 \Omega  r_+ (4 \Omega  (9 \beta +70 \Omega )+109 \mu_0
   ^2)-2 r_+^2 (2 \mu_0 ^2 \Omega  (24 \beta +209 \Omega )
   \nonumber\\
   &&
   +8 \Omega ^3 (51 \beta +58 \Omega )+47 \mu_0 ^4)+8
   \beta  \mu_0 ^4 \Omega ^2 r_+^7 (3 \mu_0 ^2+2 \Omega ^2)-2 r_+^3 (\beta  (6 \mu_0 ^4+248 \mu_0 ^2
   \Omega ^2+816 \Omega ^4)
   \nonumber\\
   &&
   +53 \mu_0 ^4 \Omega +210 \mu_0 ^2 \Omega ^3+112 \Omega ^5)+\mu_0 ^2 r_+^6 (2 \mu_0 ^4
   \Omega  (11 \beta +5 \Omega )+8 \mu_0 ^2 \Omega ^3 (4 \beta +\Omega )-192 \beta  \Omega ^5-\mu_0 ^6)
   \nonumber\\
   &&
   +r_+^4 (-2
   \mu_0 ^4 \Omega  (16 \beta +3 \Omega )-24 \mu_0 ^2 \Omega ^3 (38 \beta +3 \Omega )+64 \Omega ^5 (\Omega -21 \beta )+19 \mu_0
   ^6)
   \nonumber\\
   &&
   +r_+^5 (\beta  (6 \mu_0 ^6-4 \mu_0 ^4 \Omega ^2-704 \mu_0 ^2 \Omega ^4-384 \Omega ^6)+19 \mu_0 ^6
   \Omega +24 \mu_0 ^4 \Omega ^3+32 \mu_0 ^2 \Omega ^5)-408 \Omega ^2)
   \nonumber\\
   &&
   +\mathcal O((r-r_+)^2), 
   \end{eqnarray}
   \begin{eqnarray}
   H_{2-}&=&
   \frac{(r-r_+)}{3 r_+ (\Omega  r_++1)
   (2 \Omega  r_++1)^2 (\mu_0 ^2 r_++2 \Omega )^2 (\mu_0 ^2 r_++4 \Omega )}
   \nonumber\\
   &&
    (\mu_0 ^6 r_+^3 (12 \beta  r_++\mu_0 ^2 r_+^2+3 \mu_2^2 r_+^2-21)-8
   \Omega ^5 r_+^2 (-24 \beta  r_+ (3 \mu_0 ^2 r_+^2+4)+2 \mu_0 ^4 r_+^4-21 \mu_0 ^2
   r_+^2
   \nonumber\\
   &&
   -24 \mu_2^2 r_+^2-62)-4 \Omega ^4 r_+ (-24 \beta  r_+ (3 \mu_0 ^4
   r_+^4+12 \mu_0 ^2 r_+^2+5)+\mu_0 ^6 r_+^6+6 \mu_0 ^4 r_+^4
   \nonumber\\
   &&
   -\mu_0 ^2 r_+^2 (36 \mu_2^2
   r_+^2+125)-60 \mu_2^2 r_+^2-64)+2 \Omega ^3 (24 \beta  r_+ (\mu_0 ^6 r_+^6+12
   \mu_0 ^4 r_+^4+15 \mu_0 ^2 r_+^2+2)
   \nonumber\\
   &&
   -13 \mu_0 ^6 r_+^6+\mu_0 ^4 (24 \mu_2^2 r_+^6+22
   r_+^4)+6 \mu_0 ^2 r_+^2 (17 \mu_2^2 r_+^2+22)+12 \mu_2^2 r_+^2-8)
   \nonumber\\
   &&
   +\mu_0
   ^2 \Omega  r_+^2 (\mu_0 ^4 r_+^2 (60 \beta  r_++9 \mu_2^2 r_+^2-19)+8 \mu_0 ^2
   (9 \beta  r_++3 \mu_2^2 r_+^2-11)-6 \mu_0 ^6 r_+^4-6 \mu_2^2)
   \nonumber\\
   &&
   +2 \Omega ^2 r_+
   (\mu_0 ^6 r_+^4 (48 \beta  r_++3 \mu_2^2 r_+^2-25)+3 \mu_0 ^4 r_+^2 (60 \beta 
   r_++12 \mu_2^2 r_+^2+5)
   \nonumber\\
   &&
   +\mu_0 ^2 (72 \beta  r_++27 \mu_2^2 r_+^2-52)-2 \mu_0 ^8
   r_+^6-12 \mu_2^2)+32 \Omega ^6 r_+^3 (12 \beta  r_++7))
\cr
&&
   +\mathcal O((r-r_+)^2), 
   \end{eqnarray}
   \begin{eqnarray}
   H_2&=&\frac{\Omega  r_+
   (\Omega  (4 \mu_0 ^2 r_+^2-6 \mu_2^2 r_+^2+6)+r_+ (4 \mu_0 ^2+3 \mu_2^2)+4
   \Omega ^3 r_+^2+10 \Omega ^2 r_+)}{6 (r-r_+) (\mu_0 ^2 r_++2 \Omega )^2}
   \nonumber\\
   &&
   +
   \frac{1}{6 (\Omega
    r_++1) (\mu_0 ^2 r_++2 \Omega )^2 (\mu_0 ^2 r_++4 \Omega )}
    (6 \Omega ^4 r_+ (8 \beta  r_+ (2 \mu_0 ^2 r_+^2+3)-5 \mu_0 ^2 r_+^2+4 \mu_2^2
   r_+^2-40)
   \nonumber\\
   &&
   +3 \mu_0 ^2 \Omega  r_+^2 (4 \beta  \mu_0 ^2 r_+-\mu_2^2)-4 \Omega ^5 r_+^2
   (-24 \beta  r_++\mu_0 ^2 r_+^2+18)
   \nonumber\\
   &&
   +2 \Omega ^3 (12 \beta  r_+ (\mu_0 ^4 r_+^4+6 \mu_0
   ^2 r_+^2+2)+3 \mu_0 ^2 r_+^2 (\mu_2^2 r_+^2-40)+6 \mu_2^2
   r_+^2-56)
   \nonumber\\
   &&
   +\Omega ^2 r_+ (12 \mu_0 ^4 r_+^2 (3 \beta  r_+-5)+\mu_0 ^2 (48 \beta 
   r_++3 \mu_2^2 r_+^2-82)-12 \mu_2^2)
   \nonumber\\
   &&
   +6 \mu_0 ^6 r_+^3-16 \Omega ^6 r_+^3)+\mathcal O(r-r_+). 
   \end{eqnarray}


\end{document}